\documentclass{aa}  
\usepackage[a-2u]{pdfx}

\hypersetup{
    unicode=true,
    pdfpagelabels=true,
    pdfa=true
}

\usepackage{graphicx}
\usepackage{txfonts}
\usepackage{lipsum}
\usepackage{subcaption}         
\usepackage{lscape}             
\usepackage{placeins}           

\begin{document}

   \title{HD~210058~b and BD+37~4838~b: A precessing eccentric hot Jupiter and a misaligned substellar companion around two A-type stars}
   \titlerunning{A precessing eccentric hot Jupiter and a misaligned substellar companion around two A-type stars}

   \author{Z. Balkóová\inst{1, 2}
        \fnmsep\thanks{Corresponding author: zuzana.balkoova@asu.cas.cz}
        \and J. Žák\inst{1}
        \and M. Skarka\inst{1}
        \and M. C. Johnson\inst{3}
        \and A. W. Mann\inst{4}
        \and M. G. Barber\inst{4}
        \and P. Gajdoš\inst{5}
        \and T. G. Beatty\inst{6}
        \and M.\,A.~Kroft\inst{6}
        \and I. Ilyin\inst{7}
        \and P. Kabáth\inst{1}
        }

   \institute{Astronomical Institute of the Czech Academy of Sciences, Fričova 298, 251~65 Ondřejov, Czechia
       \and Astronomical Institute of Charles University, V Holešovičkách 2, 180~00 Prague, Czechia
       \and Department of Astronomy, The Ohio State University, 4055 McPherson Laboratory, 140 West 18th Avenue, Columbus, OH 43210, USA
       \and Department of Physics and Astronomy, The University of North Carolina at Chapel Hill, Chapel Hill, NC 27599, USA
       \and Institute of Physics, Faculty of Science, Pavol Jozef Šafárik University, Park Angelinum 9, 040 01 Košice, Slovakia
       \and Department of Astronomy, University of Wisconsin--Madison, Madison, WI 53706, USA
       \and Leibniz-Institut für Astrophysik Potsdam, An der Sternwarte 16, 14482 Potsdam, Germany
       }

   \date{Received 24 July 2026 / Accepted 27 October 2026}

  \abstract 
   {Relatively few A- and F-type stars are known to host exoplanets, and likewise few planets have measured spin-orbit angles, defined as the angle between the planet's orbital axis and the star's rotational axis. Expanding the sample of such systems is important for constraining the formation and migration mechanisms of close-in giant planets around early-type stars.}
   {We aim to measure the sky-projected spin-orbit angles of HD~210058~b (TOI-1355~b) and BD+37~4838~b (TOI-1414~b), two objects listed as candidate exoplanets, and assess the planetary nature of these systems. Both of their host stars are fast rotators.} 
   {We analyzed TESS photometry in combination with the transit time series of HARPS-N spectra. For both objects, we created cross-correlation functions from the spectra to analyze Doppler tomography and detect and model the Doppler shadow. We also used photometric observations to refine several poorly constrained photometric and spectroscopic parameters of both the host stars and their companions. When possible, we used the derived results to constrain the companion masses.}
   {We verified the planetary and substellar nature for both objects. We report HD~210058~b to be a giant hot Jupiter with a mass of $\sim10\,\mathrm{M_{Jup}}$, residing on a slightly misaligned orbit with a sky-projected spin-orbit angle $\lambda=-32_{-6}^{+4}\,\rm deg$. Archival photometry reveals a continuously changing inclination and impact parameter, indicating nodal precession with a period of $\sim860\,\rm yr$ and implying that further transit observations will be possible for only about six more years. The second target, BD+37~4838~b, orbits a hybrid $\gamma$~Dor-$\delta$~Sct pulsator on a misaligned orbit with $\lambda=56.9^{+1.4}_{-2.3}\,\rm deg$. Since we could not reliably recover its mass, we only report an upper limit of $\sim24.3\,\mathrm{M_{Jup}}$, leaving the possible planetary nature of this object unresolved.}
   {}


   \maketitle
   \nolinenumbers

\section{Introduction}
Exoplanets characterized as orbiting A- and early F-type stars are relatively rare, and their formation is not fully understood. Naively, giant-planet formation should be favored around such stars: disk mass scales with stellar mass \citep{Andrews13, Pascucci16}, and the occurrence rate of giant planets indeed increases with stellar mass up to $\approx1.9\,\mathrm{M_{\odot}}$, before rapidly decreasing for masses higher than $2.5-3.0\,\mathrm{M_{\odot}}$ \citep{Reffert15}. This is not trivial; however, because the gas-rich disk phase is also markedly shorter around more massive stars \citep{Ribas15}, this leaves less time for a core to reach runaway gas accretion before gas dispersal. That giants nonetheless remain common implies either unusually efficient core growth in these more massive disks or a contribution from disk instability \citep{Ribas15}. The situation is less clear for the close-in giant-planet population. While \cite{Zhou19} report a numerically lower hot-Jupiter occurrence rate around hotter stars than around cooler stars, they observe no statistically significant trend with stellar mass. More recently, \cite{Giacalone2025} argued that the incline toward early-type hosts may be stronger for hot Jupiters than for small close-in planets, although the relative occurrences of these populations around hot stars have not yet been firmly established.
\par
To define a “hotter” star, we adopt A-type stars as the representative class. According to the Catalog of Exoplanets\footnote{\url{https://exoplanet.eu/catalog/}}, only 25 host stars are explicitly classified as A-types as of April 2026. However, this number is likely incomplete, as stellar spectral types are not consistently reported in such catalogs. Moreover, although they correlate with effective temperature, their boundaries are not sharply defined and can overlap with adjacent classes, particularly F-type stars. To mitigate this, we instead adopt an effective temperature criterion. Using a commonly accepted lower limit of $T_{\mathrm{eff}}=7300\,\mathrm{K}$ for A-type stars, \citep{GDR3} increased the sample to 36 systems. Furthermore, since one of the target host stars analyzed in this work has a measured effective temperature slightly below this threshold, we can consider a more inclusive limit of $T_{\mathrm{eff}}=7000\,\mathrm{K}$. This yields either 45 systems or 56 systems when including objects whose upper temperature uncertainties reach $7000\,\rm K$. Even in this expanded sample, which includes stars that host multiple planetary companions, they represent fewer than $1\,\%$ of the $\sim\,6300$ confirmed exoplanets currently known, raising the question of whether this scarcity is due to intrinsic rarity or observational bias. Such a bias could be attributed to the challenges of obtaining precise radial velocity (RV) measurements for fast rotators, as well as to their relatively large radii that produce shallower planetary transits.
\par
The percentage is slightly higher ($\sim$2--6$\,\%$ depending on the set boundary) when exoplanet-hosting stars are included down to the Kraft break, i.e., the transition between slowly rotating, convective stars with effective magnetic braking \citep{Barnes_2010, Matt_2012} and rapidly rotating, radiative stars lacking such braking. \cite{Kraft1967} determined the central effective temperature for this transition at $\sim6250\,\mathrm{K}$, although modern observations, which show the expanded distribution, rather than a simple trend, shift this value to $\sim6500\,\mathrm{K}$ \citep{Beyer_2024, Wang2026}. Moreover, \cite{Spalding_2022} show that metallicity also plays a significant role in the division between convective and radiative stellar envelopes. As discussed in \cite{Beyer_2024}, the very few exceptions are fast-rotating cool stars that are either young ($< 1\, \rm Gyr$), candidate binaries, or evolved sub-giants. On the other hand, low projected rotational velocities ($v\sin{i_\star}$) in hotter stars are most naturally explained by very low (pole-on) stellar inclination angles.
\par
\cite{Winn10_obliquities} first noticed that the distribution of spin-orbit angles ($\lambda$) is much wider for planets orbiting stars above the Kraft break, whereas below the break, the $\lambda$ values remain very low, indicating orbits primarily aligned with the rotational axes of their host stars. This is not surprising, since the radiative envelopes of hot stars prevent effective tidal dissipation, resulting in theoretical realignment timescales several orders longer than in convective stars \citep{Albrecht2012}. According to the TEPCat catalog\footnote{\url{https://www.astro.keele.ac.uk/jkt/tepcat/obliquity.html}}, as of April 2026, only 45 objects orbit around stars at or above the ``newly defined'' Kraft break ($T_{\mathrm{eff}}=6500\,\mathrm{K}$) with measured projected spin-orbit angles. Since we do not know much about planet formation around hot stars, this is of particular importance.
\par
Dynamical studies can be ambiguous, and additional clues can be sought through exoplanetary atmospheres. A complete characterization therefore requires both atmospheric and dynamical tracers. Hot Jupiters around F- and A-type stars above the Kraft break are attractive targets for JWST and Ariel missions, whose transmission and, importantly, emission spectroscopy can measure C/O ratios and metallicities. Emission measurements in particular can constrain composition without requiring a precise planetary mass \citep{Guo26}. On their own, however, planetary atmospheric compositions can be degenerate tracers of initial conditions in the protoplanetary disk and/or chemical precesses in the atmospheres \citep{Oberg_2011, Penzlin_2024}, and the dynamical context required for breaking that degeneracy is often missing. \cite{Zak_2025} observe that the majority of planets with atmospheric data still lack a spin–orbit angle measurement, which may limit the interpretation of upcoming atmospheric surveys. Measuring $\lambda$ for systems such as HD~210058\,b and BD+37~4838\,b directly addresses this gap.

\section{Target selection}
Both HD~210058~b (also labeled TOI-1355~b) and BD+37~4838~b (also labeled TOI-1414~b) are listed as candidate planets orbiting around hot stars (Table~\ref{tab:stellar_params}). They were discovered by the TESS mission in 2019, and, at least to our knowledge, still remain unpublished with an in-depth investigation\footnote{During the peer-review process of this manuscript, we became aware of \citet{Watanabe26} and their analysis of TOI-1355, which is in good agreement with our results.} (although BD+37~4838 was originally identified as a transit candidate by the KELT team, designated as KELT-5). The follow-up ended up being ambiguous, but the imaging results were published in \cite{Coker_2017} with their high-resolution imaging survey, which resolved no stellar companion to BD+37~4838 within the detection limits. According to the stellar parameters reported in the ExoFOP database by FIES and/or TRES, two high-resolution echelle spectrographs, both of our targets' hosts exhibit significant rotational broadening (with $v\sin{i_\star}$ of approximately $100\, \rm km\,s^{-1}$), making these systems promising for spectroscopic follow-up and Doppler tomographic analyses.
\par
In Table~\ref{tab:stellar_params}, we also report the original A2 stellar classification of HD~210058 \citep{Cannon1918}, the F0 classification of BD+37~4838 by \citet{Tian_2023}, and its identification as a chemically peculiar star by \citet{Zhang_2023}. Additionally, we provide the \textit{Gaia} Data Release 3 (DR3) renormalized unit weight error (RUWE) values for both target stars. The RUWE is a measure of the quality of the \textit{Gaia} astrometric solution: values close to unity indicate that the observations are well described by a single-star model, while elevated values may point to unresolved multiplicity or other astrometric anomalies. Both of our targets have RUWE values close to unity, providing no indication of significant astrometric anomalies. This information plays a role in our discussion of the possible formation, migration, and evolutionary histories of companions in these systems.

\begin{table}[]
    \caption{Stellar parameters of HD~210058 and BD+37~4838.}
    \centering
    \begin{tabular}{lcc}
         \hline\hline
         \noalign{\smallskip}
         star & HD~210058 & BD+37~4838 \\
         \noalign{\smallskip}
         \hline
         \noalign{\smallskip}
         $G$ (mag) & 8.72 & 11.32 \\
         Sp. type & A2 \textsuperscript{(1)} & F0 \textsuperscript{(2)} -- Am \textsuperscript{(3)} \\
         $T_{\mathrm{eff}}\,\mathrm{(K)}$ & $9218_{-437}^{+835}$ & $7128\pm132$ \\
         $R\,\mathrm{(R_{\odot})}$ & $1.853_{-0.072}^{+0.062}$ & $1.588_{-0.075}^{+0.060}$ \\
         $M\,\mathrm{(M_{\odot})}$ & $2.34_{-0.24}^{+0.53}$ & $1.59_{-0.20}^{+0.35}$ \\
         $L\,\mathrm{(L_{\odot})}$ & $22.34_{-3.37}^{+5.91}$ & $5.86_{-0.34}^{+0.29}$ \\
         $\log{g}$ (cm\,s$^{-2}$) & $4.271_{-0.081}^{+0.092}$ & $4.238_{-0.059}^{+0.114}$ \\
         RUWE & 0.83 & 1.079 \\
         \noalign{\smallskip}
         \hline
    \end{tabular}
    \tablefoot{Magnitude and RUWE were taken from \textit{Gaia} DR3 measurements \citep{GaiaCollab}, while spectral types were given by \textsuperscript{(1)}\citet{Cannon1918}, \textsuperscript{(2)}\citet{Zhang_2023}, and \citet{Tian_2023}\textsuperscript{(3)}. All other parameters come from TESS Input Catalog \citet{Stassun19}.}
    \label{tab:stellar_params}
\end{table}

\section{Data}
\subsection{Photometry}
For HD~210058, we found and downloaded TESS photometric data from sectors 57, 77, and 84, which were, in this order, obtained in TESS $\sim$27-day slots starting on 2022 September 30, 2024 March 25, and 2024 October 01. We adopted the 200 s FFI-based light curves generated by the \texttt{QLP} pipeline \citep{Huang20a, Huang20b, Kunimoto21}, motivated by its strong performance in phase-curve extraction and mitigation of background contamination \citep{Han23}. \cite{Han23} show that the FFI light curves are broadly comparable to the 2 min \texttt{SPOC} light curves \citep{Jenkins16, Caldwell20} (at least for stars brighter than approximately 12th magnitude) while preserving transit signals and removing stellar trends.
\par
In addition to the 200 s data, we downloaded older TESS 1800-s data from the year 2019. Since such a high cadence can smear both ingress and egress in individual transit profiles and thus reduce the precision of the inferred parameters \citep{Hernandez23, Huber22}, we compensated for this by stitching together three available 1800 s cadence sectors (16, 17, 18) prior to fitting the phase curve. In this way, we combined 26 transits, each depicted by only five data points on average, into one nicely covered phase curve.
\par
For BD+37~483, the number of available TESS measurements was smaller than in the previous case. To maintain consistency with the primary data type used, we again examined the 200 s \texttt{QLP} datasets and found only two, specifically in sectors 57 and 84, corresponding to the observation slots that began on 2022 September 30 and 2024 October 01. Only one additional sector, (sector 16 from 2019 September 12) contained \texttt{QLP} data for BD+37~4838. These, however, were available only at the 30 min cadence. Fortunately, the same sector was also available as 120 s \texttt{SPOC} data, so we opted for the shorter cadence rather than the same pipeline, thereby covering the phase curve more tightly. In the end, we were left with three sectors containing 6, 8, and 8 transits, respectively.

\subsection{Spectroscopy}
Searching through the HARPS-N spectrograph archive \citep{HARPSN, HARPSN_Cosentino}, we found two nights of publicly available observations of HD~210058, obtained on 2022 July 03 and 2022 July 16, containing 37 (20 in-transit) and 36 (19 in-transit) spectra, respectively. The archive contains a single night of observations of BD+37~4838, obtained on 2015 December 17 with 21 spectra, including 8 pre-transit and 13 in-transit; the observation ended at the egress of the companion.
\par
In addition, we were provided two other datasets for BD+37~4838. The first was obtained on 2017 November 15 with the PEPSI high-resolution spectrograph \citep{Strassmeier_Pepsi} and consists of 18 spectra, ten of which were obtained during transit. The second dataset -- the only one not focused on transit observations -- was obtained throughout October 2024 with the NEID spectrograph \citep{Schwab18} and holds promise for deriving a RV curve.

\section{Analysis and results}
\subsection{Methodology}
To derive the projected orbital alignment of our two targets relative to their host stars, the system geometry must first be constrained through a set of orbital parameters, most of which can be obtained from photometric observations. However, the only available values we could find originated from ExoFOP\footnote{\url{https://exofop.ipac.caltech.edu/tess/}}, which did not provide the full set of parameters required for our analysis. We therefore performed an independent analysis of the photometric time series of both HD~210058 and BD+37~4838 prior to spectroscopic study, using data either available in the MAST archive or accessed via the \texttt{lightkurve} \citep{lightkurveDoc} tool. We then removed any visible outliers from the extracted data and proceeded to use the publicly available Python software package, \texttt{allesfitter} \citep{allesfitter-code, allesfitter-paper}, which combines several useful tools relevant for light curve and RV modeling, such as \texttt{ellc} \citep{ellc} and nested sampling implemented via \texttt{dynesty} \citep{dynesty}.
\par
We investigated the Rossiter-McLaughlin (RM) effect \citep{Rossiter1924, McLaughlin1924} through Doppler tomography, which is a technique that resolves the transiting companion's trace -- known as the Doppler shadow --  as it moves across the stellar disk. For this, a series of cross-correlation functions (CCFs) derived from spectroscopic observations are required. Although pipeline-processed CCFs from HARPS-N are publicly available, their velocity range of $\langle-20, +20\rangle\,\rm km\,s^{-1}$ is insufficient for our two hot, rapidly rotating stars. For this, we chose to process the 1D-type spectra using a short self-written script for \texttt{iSpec} \citep{BCuaresma2014}, which automates cross-correlation, trimming, and extraction of CCFs at a given wavelength range. We restricted this wavelength range to $4500-5800\,\AA$ to minimize telluric contamination. As a template we used an unbroadened synthetic stellar spectrum, computed at a resolving power of 115000 consistent with HARPS-N, and sampled with a velocity increment of $0.25\, \rm km \, s^{-1}$. This was generated using \texttt{iSpec} with the \texttt{SPECTRUM} radiative transfer code \citep{SPECTRUM}. The adopted stellar parameters correspond to those listed in Table~\ref{tab:stellar_params}. Elemental abundances were adopted from \cite{Grevesse1998}, using the Gaia-ESO Survey (GES) line list, and the model atmospheres were selected from the MARCS GES-APOGEE grid \citep{Plez2008}. The same CCF extraction methodology was used by \cite{Balkoova26}.
\par
For each dataset, we compared the CCFs obtained during transit with an average out-of-transit CCF constructed from all exposures taken before and after transit. The average out-of-transit CCF represents the undistorted stellar line profile; subtracting it from each individual in-transit CCF reveals the local distortion caused by the planet occulting the rotating stellar surface. These residual CCFs trace the Doppler shadow as it moves across the stellar line profile during the transit. We modeled the Doppler shadow using the Python package \texttt{tracit}, introduced and described by \cite{Knudstrup2022}. The implemented methods aim at generating synthetic CCFs by modeling the stellar surface and distortion induced by the transiting planet. The model CCFs are then compared with the observed CCFs within a Markov chain Monte Carlo (MCMC) framework to infer the system parameters, including the projected spin-orbit angle ($\lambda$) -- the primary parameter of interest.

\subsection{HD~210058}
\subsubsection{Photometry}
\begin{figure}
    \centering
    \includegraphics[width=\columnwidth]{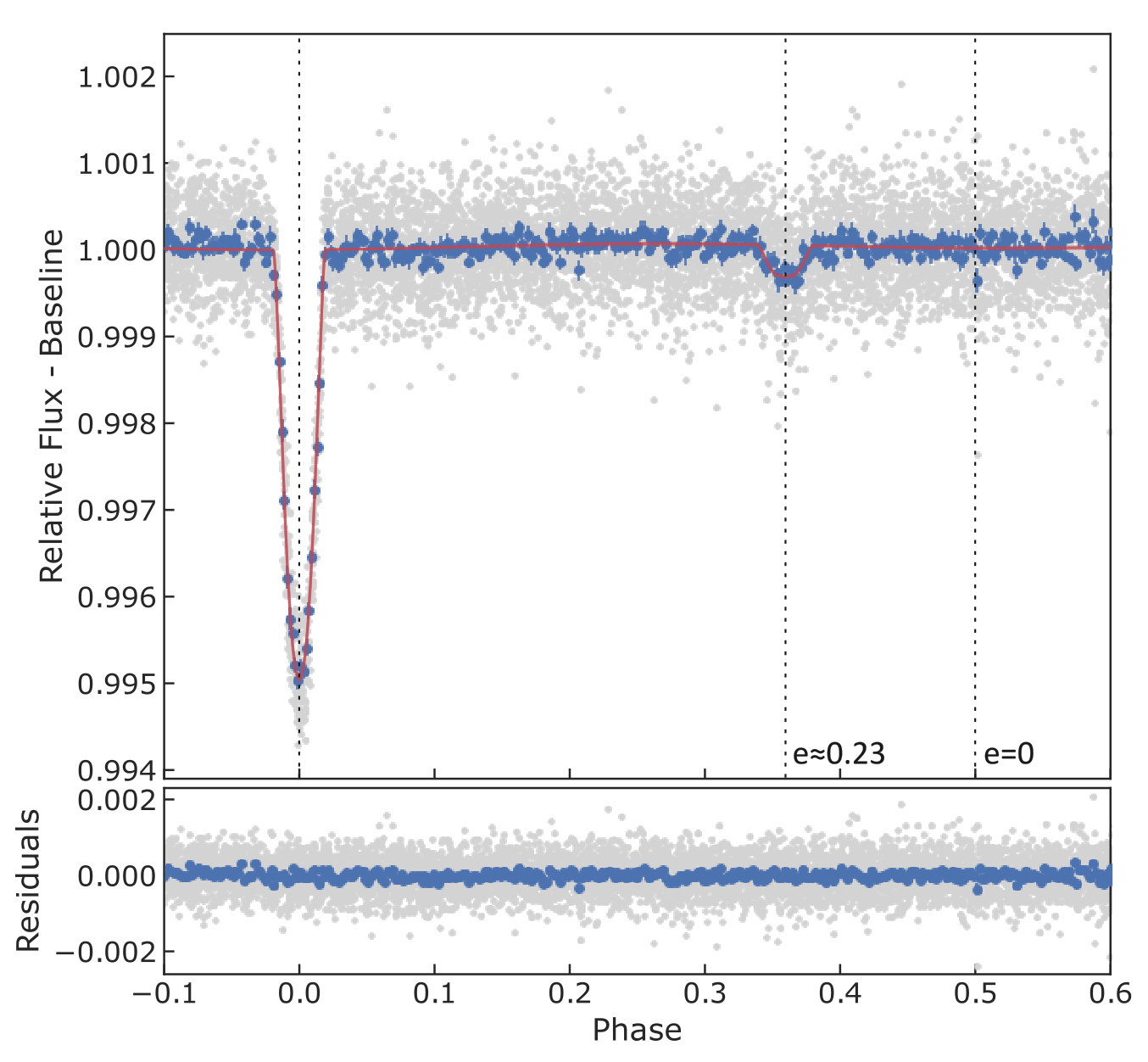}
    \caption{Folded phase curve of HD~210058 from TESS measurements (sector 84). Individual photometric measurements are shown as gray dots, binned as blue dots. The red line shows the fit to the data. The vertical dashed lines depict (from left to right) the transit, the occultation dip for the current eccentricity, and the phase at which the occultation dip would occur for a circular orbit.}
    \label{fig:TOI-1355_phaseCurve}
\end{figure}

Our initial attempt to infer the basic photometric parameters involved sampling a combined set of measurements from three sectors, which together cover 33 transits. However, upon zooming-in on the folded phase curve, we noticed inconsistent transit dip shapes, which could primarily be attributed to transits from sector 57. Therefore, we ran the sampler on each TESS sector individually (Fig.~\ref{fig:TOI-1355_phaseCurve}), revealing a continuous change in impact parameter and transit duration (Table~\ref{tab:toi-1355_photo_params}). Such behavior implies gradual variation in the inclination of the companion's orbital plane relative to the line of sight, which shifts the projected transit chord across the stellar disk. This indicates nodal precession.
\par
We verified that these varying parameters might indeed be physical by using \texttt{SPOC}-processed data from the same sectors. Although \texttt{SPOC}-processed sector 84 was unavailable, we could still use data from sectors 57 and 77 and obtain results within $1\sigma$ with those from the \texttt{QLP} data. This demonstrates that both the \texttt{QLP} and \texttt{SPOC} pipelines produce equivalent light curves and, consequently, equivalent results.
\par
\begin{figure*}
    \centering
    \includegraphics[width=0.8\linewidth]{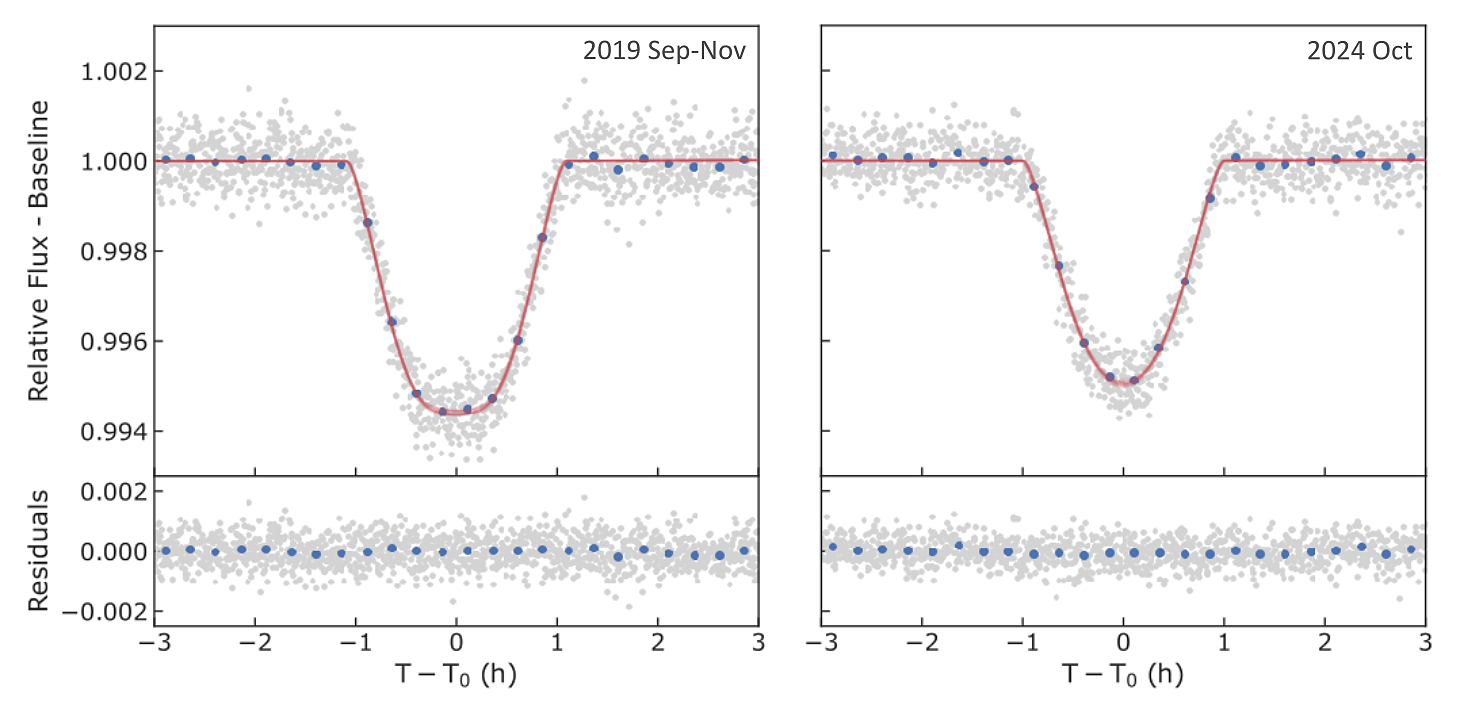}
    \caption{Folded and binned transit profiles of two datasets of HD~210058 obtained at the most widely separated epochs. Left panel: Data from 2019. Right panel: Data from 2024, in which the transits occur already in an almost grazing geometry.}
    \label{fig:precession}
\end{figure*}

To extend the time span of the photometric measurements, we additionally fitted the stitched 1800 s cadence data and added the inferred results to those of the later-observed sectors (Table~\ref{tab:toi-1355_photo_params}), further supporting the continuously evolving transit geometry. The greatest difference, visible through a comparison of the oldest and newest datasets, can be seen in Fig.~\ref{fig:precession}. In addition, all sectors consistently yield $e\sim0.22$. This is visible in the occultation dip (Fig.~\ref{fig:TOI-1355_phaseCurve}) occurring around phase 0.37, since for a circular orbit, its time difference from transit would be, if visible, exactly half a phase. With HD~210058 being a close-in giant planet, this eccentricity would be somewhat unusual. Adopting a broad definition of hot Jupiters ($P<10\,\mathrm{d}$ and $0.3<M_\mathrm{p}<13\,M_\mathrm{{Jup}}$), we find from the Catalog of Exoplanets that only about $8\,\%$ of known systems exhibit eccentricities above $e=0.1$, roughly half of which have $e<0.2$. These objects belong to the relatively small population of eccentric hot Jupiters, often regarded as potential evidence for high-eccentricity migration scenarios \citep[e.g.][]{Fabrycky2007, Giacalone2017}, in which planets are first excited onto eccentric orbits and subsequently circularized through tidal interactions.

\begin{table*}[]
    \caption{Inferred transit parameters of HD~210058~b.}
    \centering
    \begin{tabular}{lcccc}
        \hline\hline
        \noalign{\smallskip}
        Date & 2019 Sep-Nov & 2022 Sep 30 & 2024 Mar 25 & 2024 Oct 01 \\
        \noalign{\smallskip}
        \hline
        \noalign{\smallskip}
        No. transits & 26 & 13 & 9 & 11 \\
        Cadence (s) & 1800 & 200 & 200 & 200 \\
        $T_0$ (BJD-2457000) & $1777.13143\pm0.00017$ & $2868.77405\pm0.00014$ & $3409.16991\pm0.00021$ & $3597.98274\pm0.00015$ \\
        $P$ (d) & $2.170276\pm0.000015$ & $2.170230\pm0.000031$ & $2.170225\pm0.000040$ & $2.170269\pm0.000038$ \\
        $a/R_\star$ & $4.27_{-0.18}^{+0.13}$ & $4.740\pm0.091$ & $4.36_{-0.11}^{+0.13}$ & $4.43_{-0.14}^{+0.15}$ \\
        $R/R_\star$ & $0.0803_{-0.0012}^{+0.0009}$ & $0.0792\pm0.0011$ & $0.0777_{-0.0021}^{+0.0025}$ & $0.0800_{-0.0028}^{+0.0043}$ \\
        $R$ (R\textsubscript{Jup}) & $1.443\pm0.056$ & $1.425\pm0.055$ & $1.400_{-0.063}^{+0.071}$ & $1.444_{-0.075}^{+0.093}$ \\
        $a$ (au) & $0.0366\pm0.0019$ & $0.0408\pm0.0017$ & $0.0376_{-0.0017}^{+0.0018}$ & $0.0381\pm0.0018$ \\
        $e$ & $0.236_{-0.015}^{+0.017}$ & $0.2293_{-0.0044}^{+0.0052}$ & $0.2212_{-0.0026}^{+0.0029}$ & $0.2186_{-0.0022}^{+0.0030}$ \\
        $\omega$ (deg) & $203.2_{-10.}^{+7.7}$ & $197.2_{-3.9}^{+3.6}$ & $180.9_{-5.1}^{+6.5}$ & $184.1\pm3.8$ \\
        $i$ (deg) & $78.38_{-0.98}^{+0.68}$ & $79.11_{-0.38}^{+0.35}$ & $76.98_{-0.58}^{+0.67}$ & $77.22\pm0.62$ \\
        $b$ & $0.8972_{-0.0069}^{+0.0061}$ & $0.9103\pm0.0046$ & $0.9380_{-0.0070}^{+0.0083}$ & $0.9458_{-0.0095}^{+0.0120}$ \\
        $T_{\mathrm{total}}$ (h) & $2.571_{-0.026}^{+0.028}$ & $2.162\pm0.015$ & $2.028\pm0.022$ & $1.986_{-0.020}^{+0.021}$ \\
        $T_{\mathrm{full}}$ (h) & $0.862\pm0.098$ & $0.515_{-0.093}^{+0.084}$ & $0.177_{-0.091}^{+0.067}$ & $0.180_{-0.087}^{+0.033}$ \\
        $q_{\mathrm{1}}$ & $0.242_{-0.091}^{+0.082}$ & $0.179_{-0.074}^{+0.079}$ & $0.120_{-0.070}^{+0.086}$ & $0.152_{-0.073}^{+0.085}$ \\
        $q_{\mathrm{2}}$ & $0.271_{-0.086}^{+0.075}$ & $0.243\pm0.087$ & $0.252\pm0.094$ & $0.245\pm0.089$ \\
        \noalign{\smallskip}
        \hline
    \end{tabular}
    \label{tab:toi-1355_photo_params}
\end{table*}

\subsubsection{Doppler tomography}
\begin{figure}
    \centering
    \includegraphics[width=\columnwidth]{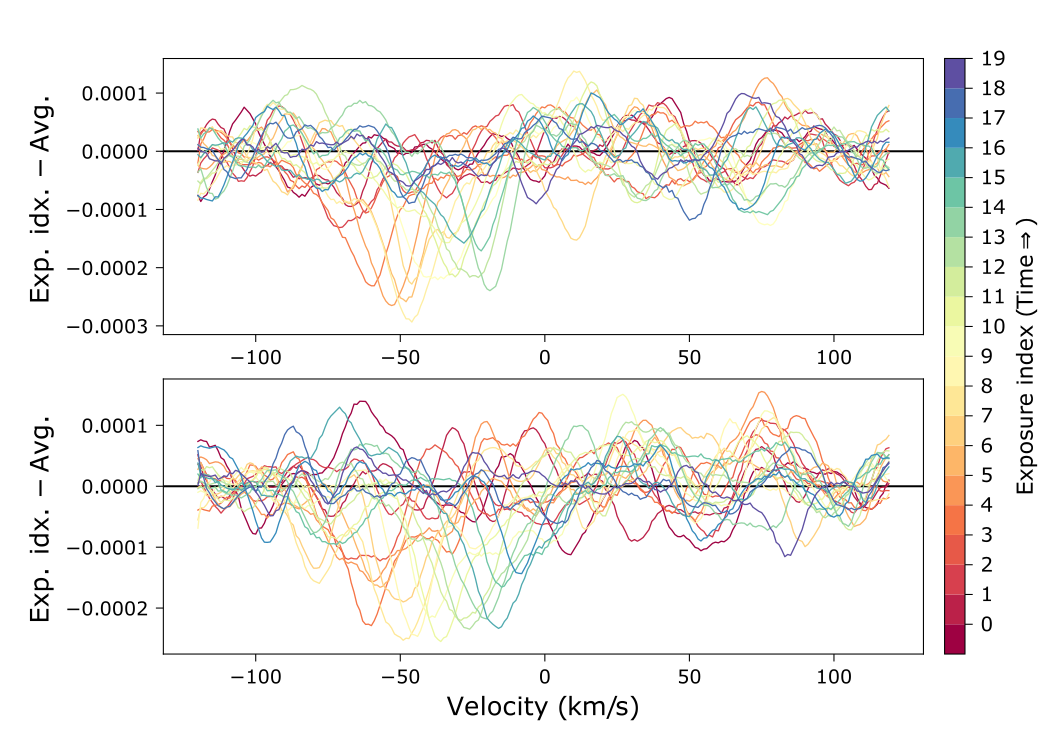}
    \caption{Residual CCFs of HD~210058 from the two nights of HARPS-N observations, obtained by subtracting the average out-of-transit CCF from each individual CCF. The color map represents the numbering of transit spectra (ingress in red, egress in blue).}
    \label{fig:toi-1355_CCFresiduals}
\end{figure}

Using the obtained photometric parameters to construct a transit model, the companion's trace is visible progressing through line distortions (Fig.~\ref{fig:toi-1355_CCFresiduals}). It only sticks to negative velocities, which means that the companion does not cross the redshifted part of the stellar surface. Since the orbit is precessing, we opted for the modeling values derived from fitting the 57th TESS sector measurements (second column of Table~\ref{tab:toi-1355_photo_params}), corresponding to the year 2022. These most closely represent the spectroscopic data, which are only three months older.
\par
We, moreover, inferred five relevant spectroscopic parameters: the projected rotational velocity ($v\sin i_\star$), projected spin-orbit angle ($\lambda$), micro- and macroturbulent velocity, and systemic velocity ($\gamma$). In particular, we focused on $v\sin i_\star$ and $\lambda$, for which we adopted wide uniform priors (Table~\ref{tab:DSs_priors_posteriors}), since neither parameter had been previously measured with sufficient precision. Next, we adopted a Gaussian prior of $\mathcal{N}(-21, 10.6)\,\rm km\,s^{-1}$ on the $\gamma$ velocity, which is based on an initial  \texttt{iSpec} estimate obtained by comparing the barycentrically corrected HARPS-N spectra with a synthetic spectrum. The prior standard deviation is very broad here due to rotational broadening,which limits the precision of velocity-shift measurements in the line profiles. Consequently, the sampler converged only to solutions in which the Doppler shadow was visibly misaligned with the model. We therefore inspected the velocity alignment of the residual CCF maps directly and chose a more restrictive prior of $\mathcal{N}(-24.7, 1)\,\rm km\,s^{-1}$. Similarly, determining the turbulence velocity, particularly the macroturbulence velocity, is also challenging for fast rotators. According to \citep{Doyle2014}, who derived a calibration between $T_{\mathrm{eff}}$, $v_{\mathrm{mac}}$, and $log_{10}\,g$ based on a sample of primarily convective F and G stars, stellar line blending becomes an issue already at $v\sin i_\star$ $\sim12\,\rm km\,s^{-1}$. Since our 9000-kelvin target's projected rotational velocity largely exceeds this threshold, we estimate the prior distribution of $v_{\mathrm{mac}}$ to be slightly elevated from the "typical" value. At the same time, we set the mean prior value of $v_{\mathrm{mic}}$ to $2\,\rm km\,s^{-1}$, since microturbulences in A and F stars are not expected to exceed $\sim4\pm1\,\rm km\,s^{-1}$ \citep{Gebran2014}.
\par
\begin{figure*}
    \centering
    \includegraphics[width=\linewidth]{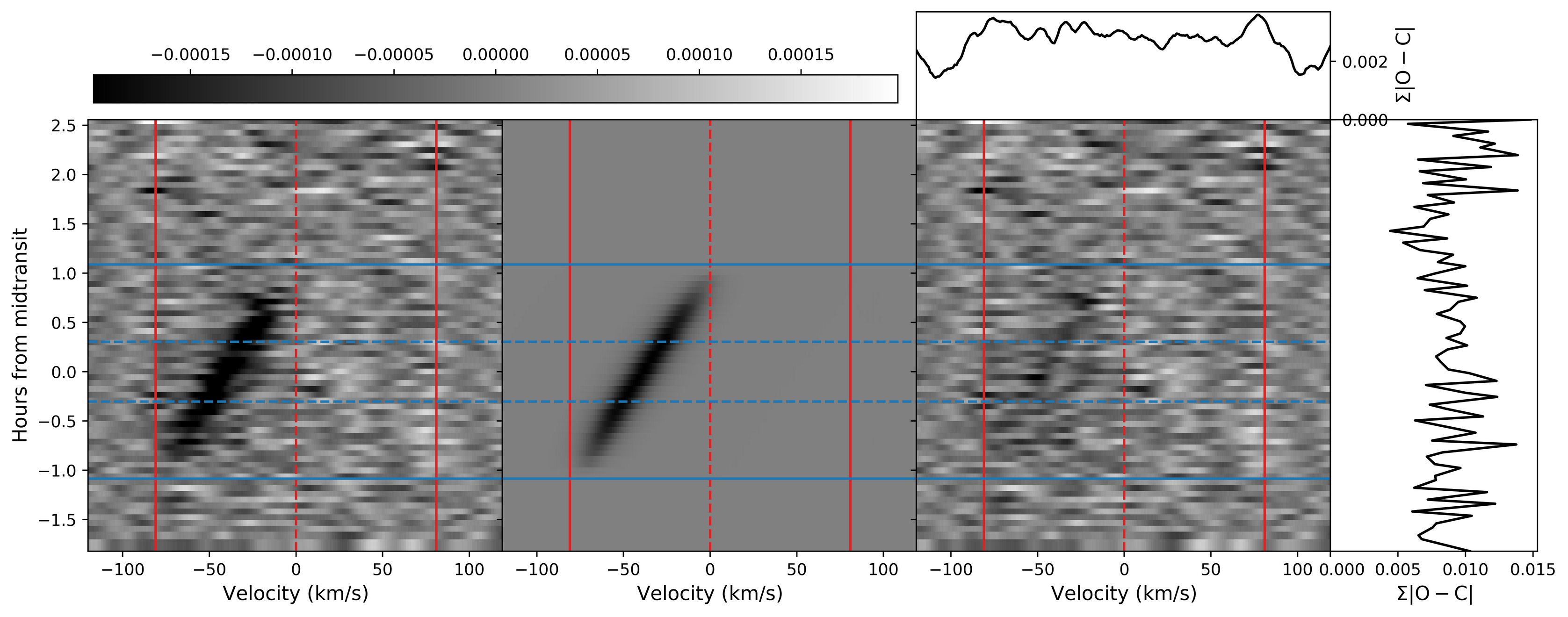}
    \caption{Doppler tomography of HD~210058 combining observations from the two nights. \textit{Left}: Observed line-profile residuals. \textit{Middle}: Best fitting Doppler-shadow model. \textit{Right}: Residuals after model subtraction. Top and far-right panels: Sums of absolute residuals along the time and velocity axes, respectively. The vertical red lines mark the approximate limits of the stellar rotational profile, while the dashed blue lines indicate the extent of the transit.}
    \label{fig:toi-1355_DT}
\end{figure*}

The independently fitted datasets from the two observing nights yield mutually consistent results for $v\sin i_\star$ and $\lambda$ within uncertainties (Table~\ref{tab:DSs_priors_posteriors}), indicating a moderately misaligned prograde configuration. This interpretation is qualitatively consistent with the observed evolution of the orbital inclination due to nodal precession, since no such change would occur for a well-aligned orbit. We obtain projected stellar obliquities of $-37_{-7}^{+8}\,\rm deg$ and $-30\pm5\,\rm deg$ for the first and second night, respectively. Night~2 provides a somewhat tighter constraint on $\lambda$, which may be partially due to its slightly higher average signal-to-noise ratio {at $550$\,nm}, exceeding that of night~1 by 5.55. Finally, a joint fit of both nights in Fig.~\ref{fig:toi-1355_DT} reveals $\lambda=-32_{-6}^{+4}\,\rm deg$ and $v\sin i_\star=81_{-11}^{+9}\,\rm km\,s^{-1}$.
\par
We note that the resulting uncertainties on the $\gamma$ velocity are extremely small and reflect only the formal precision of the tomographic fit. The synthetic template used during CCF extraction had much narrower spectral lines than the rotationally broadened observed spectra, which can affect the detailed CCF shape and centroid. For this reason, we used $\gamma$ as a velocity zero-point parameter of the Doppler-tomographic model and do not interpret its formal uncertainty as the absolute uncertainty of the stellar systemic velocity.
\par

\begin{table*}[]
    \begin{center}
    \caption{Priors and results from fitting the Doppler shadows of HD~210058 and BD+37~4838.}
        \begin{tabular}{lcccccc}
        \hline\hline
        \noalign{\smallskip}
        Parameter & Prior & \multicolumn{3}{c}{HD~210058} & \multicolumn{2}{c}{BD+37~4838} \\ 
        & & Night 1 & Night 2 & Joint & HARPS-N & PEPSI \\
        \noalign{\smallskip}
        \hline
        \noalign{\smallskip}
        Date && 2022 Jul 03 & 2022 Jul 16 && 2015 Dec 17 & 2017 Nov 15\\
        No. spectra && 37 & 36 & 73 & 21 & 18 \\
        No. transit spectra && 20 & 19 & 39 & 13 & 10\\
        Avg. S/N\textsubscript{$550\, \rm nm\,pix^{-1}$} && 80.9 & 86.5 & 83.7 & 27.6 & 53.0 \\
        Exp. time [s] && 400 & 400 & 400 & 900 & 1200 \\
        \noalign{\smallskip}
        \hline
        \noalign{\smallskip}
        $v\sin i_\star\,\rm (km\,s^{-1})$ & $\mathcal{U}(0, 150)$ & $75_{-13}^{+12}$ & $85_{-13}^{+12}$ & $81_{-11}^{+9}$ & $81\pm4$ & $94_{-3}^{+4}$  \\
        $\lambda$ (deg) & $\mathcal{U}(-180, 180)$ & $-37_{-7}^{+8}$ & $-30\pm5$ & $-32_{-6}^{+4}$ & $56.9_{-2.2}^{+1.4}$ & $43.6_{-1.0}^{+2.4}$ \\
        $v_{\mathrm{mac}}\,\rm (km\,s^{-1})$ & $\mathcal{N}(9, 5)$ & $14\pm3$ & $14\pm3$ & $16_{-2}^{+3}$ & $14.9_{-1.4}^{+1.3}$ & $13.2_{-0.6}^{+0.7}$ \\
        $v_{\mathrm{mic}}\,\rm (km\,s^{-1})$ & $\mathcal{N}(2, 1)$ & $2.1\pm1.0$ & $2.1\pm1.0$ & $2.2\pm1$ & $2.7_{-1.0}^{+0.9}$ & $1.6_{-0.9}^{+0.7}$ \\
        $\gamma\,\rm (m\,s^{-1})$ & $\mathcal{N}$* & $-23734\pm7$ & $-23735_{-8}^{+7}$ & $-23736_{-3}^{+4}$ & $257_{-137}^{+173}$ & $-750_{-87}^{+76}$ \\
        \noalign{\smallskip}
        \hline
        \end{tabular}
    \label{tab:DSs_priors_posteriors}
    \tablefoot{The posterior values are given below the horizontal rule after the exposure time parameter. \\ *The Gaussian priors were set to $\mathcal{N}(-24700, 1000)\,\rm m\,s^{-1}$ and $\mathcal{N}(-1390, 6900)\,\rm m\,s^{-1}$ for HD~210058 and BD+37~4838, respectively.}
    \end{center}
\end{table*}

\subsubsection{Nodal precession}
Generally, the precession of an orbiting planet's node is attributed to two primary mechanisms. The first is driven by gravitational perturbations by additional bodies in the system, such as other planets \citep{Bailey20}. The second mechanism is associated with the nonspherical gravitational potential of a rapidly rotating host star, whose rotationally induced oblateness gives rise to a nonzero stellar quadrupole moment \citep{Stephan_2023}. Both these effects induce secular variations in the orbital elements and lead to a precession of the orbital plane. From the observer's point of view, this can be reflected in changes in planetary inclination and, hence, the impact parameter, thereby leading to a gradual change in transit duration, shape, and, eventually, depth. In fact, this phenomenon also explains why some transiting planets can stop transiting and non-transiting planets can start. Generally, the transit probability is simply a function of the ratio of the host’s radius divided by the orbital semimajor axis of the exoplanet, $a/R_\star$. Eccentricity may alter this likelihood. However, \cite{Stephan_2023} points out that nodal precession makes it additionally a function of the precession rate, thereby increasing it by a factor of a few to $\sim10\,\%$.
\par
\begin{figure}[t]
    \centering
    \includegraphics[width=\columnwidth]{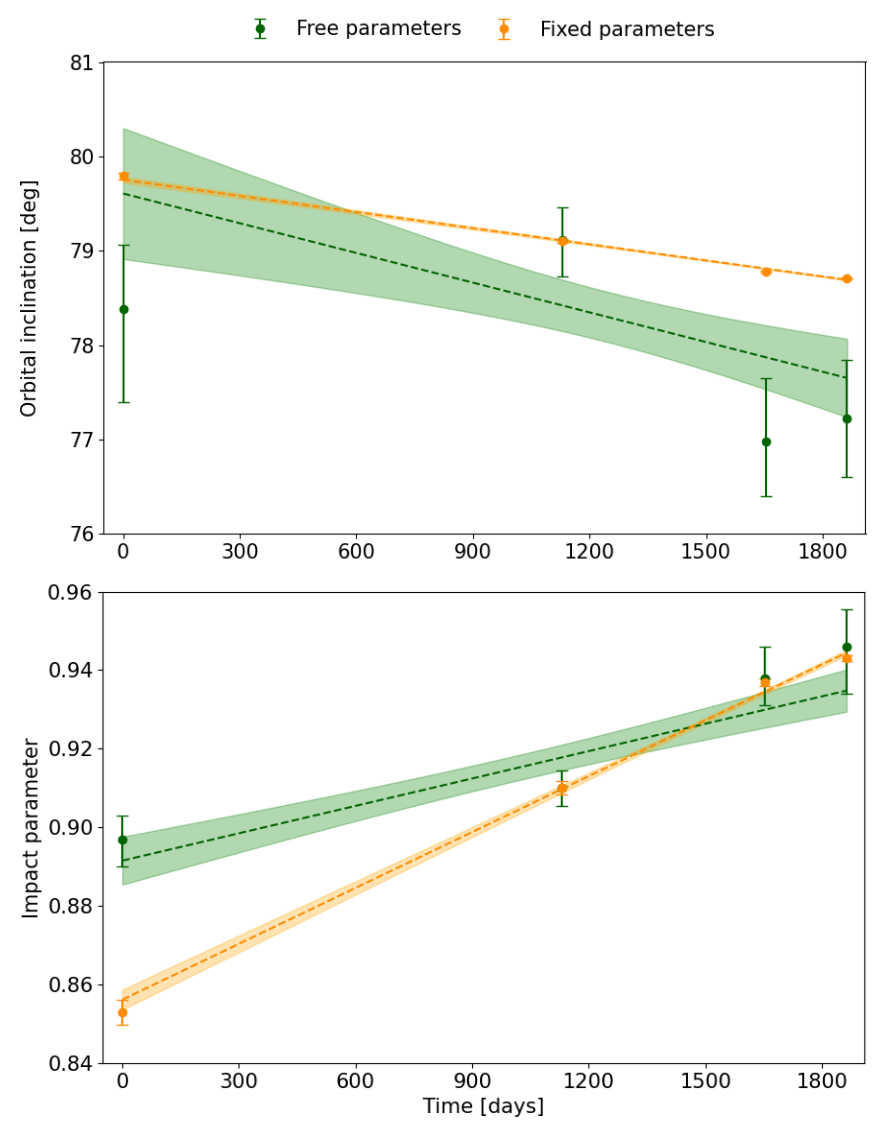}
    \caption{Evolution of the orbital inclination and impact parameter of HD 210058. The green data points show results with all transit parameters left free. The orange points represent attempts in which only the impact parameter $b$ was allowed to vary, while all other transit parameters were held fixed. Linear fits to both datasets are shown, based on results obtained from four TESS sectors. Time is relative to the central date from the first of the four analyzed datasets.}
    \label{fig:toi-1355_changes}
\end{figure}

From the results in Table~\ref{tab:toi-1355_photo_params}, we notice that the orbiting body (a term used temporarily here; the object's mass is discussed later) slowly drifts away from our view of the stellar disk, as shown by an increasing impact parameter and decreasing transit duration. We would also expect the inclination toward the observer to decrease, although this trend is not clearly seen. From the formula,
\begin{equation}
    i_{\mathrm{orb}}=\arccos{\left(\frac{bR_\star}{a}\frac{1+e\sin\omega}{1-e^2}\right)},
    \label{eq:incl}
\end{equation}
we see that inclination is directly tied to the stellar radius ($R_\star$), the semimajor axis ($a$), as well as the eccentricity and argument of periapsis ($e$ and $\omega$, respectively). During the fitting, these parameters were left free for all datasets. To investigate how the inclination would change if the majority of the information were given, we additionally fit only for $b$, keeping the remaining right-side parameters fixed at the values inferred from the 2022 data. Following this approach (Fig.~\ref{fig:toi-1355_changes}), we conclude that the orbital inclination and impact parameters might undergo changes of $di/dt=-0.2079\pm0.0068\,\rm deg\,yr^{-1}$ and $db/dt=0.0173\pm0.0006\,\rm yr^{-1}$. We justify the choice of linear trend with the assumption that our data cover only a fraction of the complete precession cycle.
\par
Consequently, the transit duration is given by \cite{Winn10} as
\begin{equation}
    T_{\mathrm{tot, full}}=\frac{P}{\pi}\sin^{-1}\!\left[\frac{R_\star}{a}\,\frac{\sqrt{(1\pm k)^2-b^2}}{\sin i_{\mathrm{orb}}}\right]\cdot\frac{\sqrt{1-e^2}}{1+e\sin\omega},
    \label{eq:transit-time}
\end{equation}
where $k$ represents the ratio of the planetary to the stellar radius. Using Eqs. (\ref{eq:incl}) and (\ref{eq:transit-time}), we can predict when the companion of HD~210058 will stop transiting by setting $T_{\mathrm{tot}}=0$. For simplicity, we again assume that the impact parameter is the only varying parameter. It must reach 1.079 for $T_{\mathrm{tot}}$ to become zero. According to the fitted linear trend in the lower panel of Fig.~\ref{fig:toi-1355_changes} -- where day zero corresponds to 20 October 2019 -- it will not be possible to observe the transits $4740_{-246}^{+213}$ days from this date. More broadly, this corresponds to October 2032, with the uncertainty spanning from May 2032 to February 2033. 
\par

At present, the list of exoplanets exhibiting transit duration variation (TDV) is not too extensive (\ref{tab:nodal-precessiong_objects}). The first object of this type was KOI-13~b, orbiting an oblate hot ($\sim8000\,\rm K$) star with a $v\sin{i_\star}$ of $65\pm10\,\rm km\,s^{-1}$ \citep{Barnes11} and a rotation period of $25.43\,\rm h$ (\cite{Szabo12}). By fitting KOI-13's consecutive light curves with transit shape models, \cite{Szabo12} previously revealed the planet's TDV and reported that both the orbital plane and the stellar spin axis indicate a nodal precession period of approximately 500 years. One of the most studied exoplanets orbiting fast rotators, WASP-33~b, is also well known for its clear signs of nodal precession. Several studies \citep{Johnson_2015, Watanabe_2020, Stephan22} have estimated its period to be $\sim900-1400$ years. \cite{Watanabe_2020} also points out that the planet transits its host star only for $\approx20\,\%$ of this period. Another case is the planet KELT-9~b, for which \cite{Stephan22} estimate a nodal precession period of around 890 years, while predicting that both KELT-9~b and WASP-33~b will cease transiting around $2074_{-10}^{+12}\,\rm CE$ and $2090_{-10}^{+17}\,\rm CE$, respectively. Similarly, the planet TOI-1518~b began transiting two decades ago and will continue to do so for approximately two hundred years \citep{Watanabe24}. We are thus in a "lucky window" for observing the varying orbital behavior of these planets.
\par
A combination of gravity darkening and nodal precession was also observed in the PTFO~8-8695 system \citep{Barnes13}. However, in this case, whether an extremely shortly orbiting planet is indeed present \citep[with $P=0.4484\,\rm d$,][]{vanEyken_2012} or whether the system is instead a binary star remains unclear \citep[e.g.][]{Bouma_2020, Ciardi_2015}. Being not able to compute the mass, HD~210058~b would fall into the same problematic category. Nevertheless, while we are not certain about the mechanism driving its nodal precession, we can estimate it. Nodal precession induced by the stellar quadrupole moment ($J_2$) \citep{Iorio11, Stephan_2023} for an eccentric companion can be determined via
\begin{equation}
    \frac{\mathrm{d}\Omega}{\mathrm{d}t} = -\frac{3\pi J_2 R_\star^2}{Pa^2(1-e^2)^2}\cos{I_{\mathrm{eq}}},
\end{equation}
where, assuming a directly equator-on view of the star (the elementary assumption here)\footnote{for $i_\star\neq90\,\rm deg$, see \cite{Stephan_2023}, Eqs. (6) and (7)}, $\cos{I_{\mathrm{eq}}}\equiv\cos{\psi}$ given by the relation between the projected and true stellar obliquity,
\begin{equation}
    \cos{\psi}=\cos{i_\star}\cos{i_{\mathrm{orb}}}+\sin{i_\star}\sin{i_{\mathrm{orb}}}\cos{\lambda}.
    \label{eq_psi}
\end{equation}
With this, we also set 
\begin{equation}
    \frac{\mathrm{d}I_{\mathrm{eq}}}{\mathrm{d}t}=0.
\end{equation}
The parameter $J_2$ is the stellar quadruple moment,
\begin{equation}
    J_2 \approx \frac{1}{3} k_2\frac{\omega_\star^2 R_\star^3}{G M_\star}
\end{equation}
\citep{Spalding20}, where $k_2$ is the Love number and $\omega_\star$ is the star's angular velocity, for which the assumption of the rotational period becomes
\begin{equation}
    P_\mathrm{rot}\leq\frac{2\pi R_\star}{v_{\mathrm{eq}}},
    \label{eq_prot}
\end{equation}
using $v_{\mathrm{eq}}\equiv v\sin{i_\star} = 81\,\rm km\,s^{-1}$.
As a first-order consistency check, we estimated the nodal precession rate under the simplifying assumption of $i_\star=90\,\rm deg$, adopting a Love number of $k_2=0.012$, representative of A-type stars \citep{Best22}. Under these conditions, the rotationally induced quadrupole moment $J_2=1.091\times10^{-4}$ results in a nodal precession rate of $\dot{\Omega}=-0.42\,\rm deg\,yr^{-1}$. For HD~210058~b, the ascending node of orbit,
\begin{equation}
    \tan\Omega=-\sin\lambda\tan{i_\mathrm{orb}},
\end{equation}
including uncertainties, yields $\Omega=72.27_{-1.72}^{+3.27}\,\rm deg$ for the first night and $\Omega=68.95_{-1.89}^{+2.71}\,\rm deg$ for the second night. However, since there are only 13 days between the two of our available spectroscopic transit measurements, we cannot observationally compare the magnitude of change in $\lambda$ and $\Omega$.
On the other hand, projecting this nodal precession onto the observable transit geometry as
\begin{equation}
    \cos{i_\mathrm{orb}}=\sin{I_\mathrm{eq}}\cos{\Omega}
\end{equation}
yields a change in the planetary orbital inclination of $\dot{i_\mathrm{orb}} = -0.2078\,\rm deg\,yr^{-1}$, consistent with the value inferred from the photometric analysis, $\dot{i_\mathrm{orb}}=-0.2079\,\rm deg\,yr^{-1}$. This makes the precession period $\approx860$ years long. Although different combinations of both $i_\star$ and $k_2$ could, in principle, reproduce similar precession rates, this approximation suffices to show that the observed inclination variation is of the order of magnitude expected from nodal precession driven by the stellar quadrupole moment. Precession induced by neighboring planetary bodies cannot be ruled out as an additional contributor, as both effects can drive orbital precession \citep{Escude_2002}. However, in the absence of evidence for additional companions in this system, we cannot quantify this contribution and therefore do not consider it to be significant.
\par
We note that in Table~\ref{tab:toi-1355_photo_params}, the fitted values of the argument of periastron ($\omega$) also suggests temporal evolution. This behavior is qualitatively consistent with apsidal precession (the rotation of the line of apsides within the orbital plane), which can be driven by, namely, general relativistic effects, tidal and rotational distortions of the star and planet, as well as gravitational perturbations from additional companions \citep{Ragozzine_2009, Escude_2002}. However, we also observe a very mild decreasing trend in the eccentricity ($e$), which is unexpected under pure apsidal precession, where it should remain approximately constant. Unfortunately, the limited number of epochs and the strong correlations between $\omega$ and $e$ prevent us from drawing a firm conclusion as to whether the precession of periapsis actually occurs and what the magnitude of such a change might be. Moreover, in our fitting setup, $\omega$ and $e$ were not sampled directly but via $\sqrt{e}\cos\omega$ and $\sqrt{e}\sin\omega$. Consequently, imposing a constant eccentricity while allowing $\omega$ to vary freely was not straightforward within this parameterization. The apparent decrease in $e$ should therefore be interpreted with caution.

\subsection{BD+37~4838}
\subsubsection{Photometry}
\begin{figure}
    \centering
    \includegraphics[width=\columnwidth]{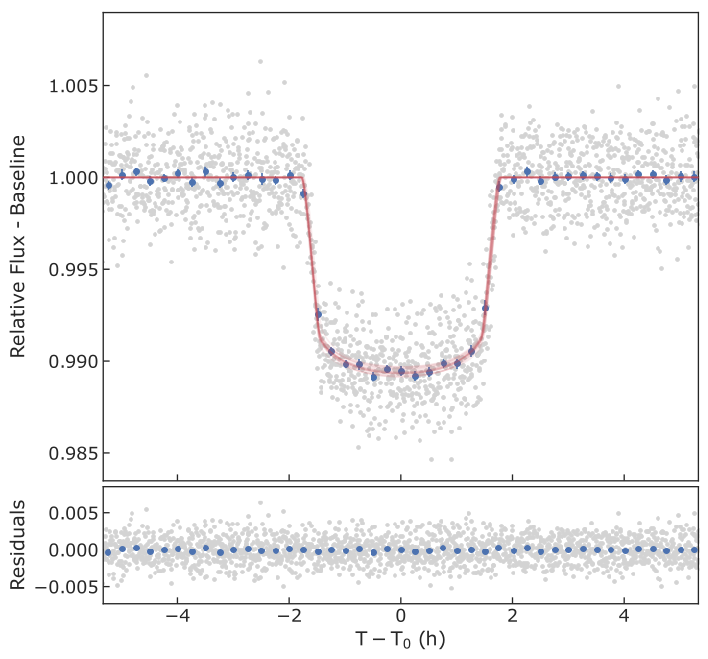}
    \caption{Folded (gray dots) and binned (blue dots) transit profiles of BD+37~4838 (sector 16). The red lines show the fitted converging transit models.}
    \label{fig:toi-1414_phasezoom}
\end{figure}

\begin{figure*}
    \centering
    \includegraphics[width=0.9\linewidth]{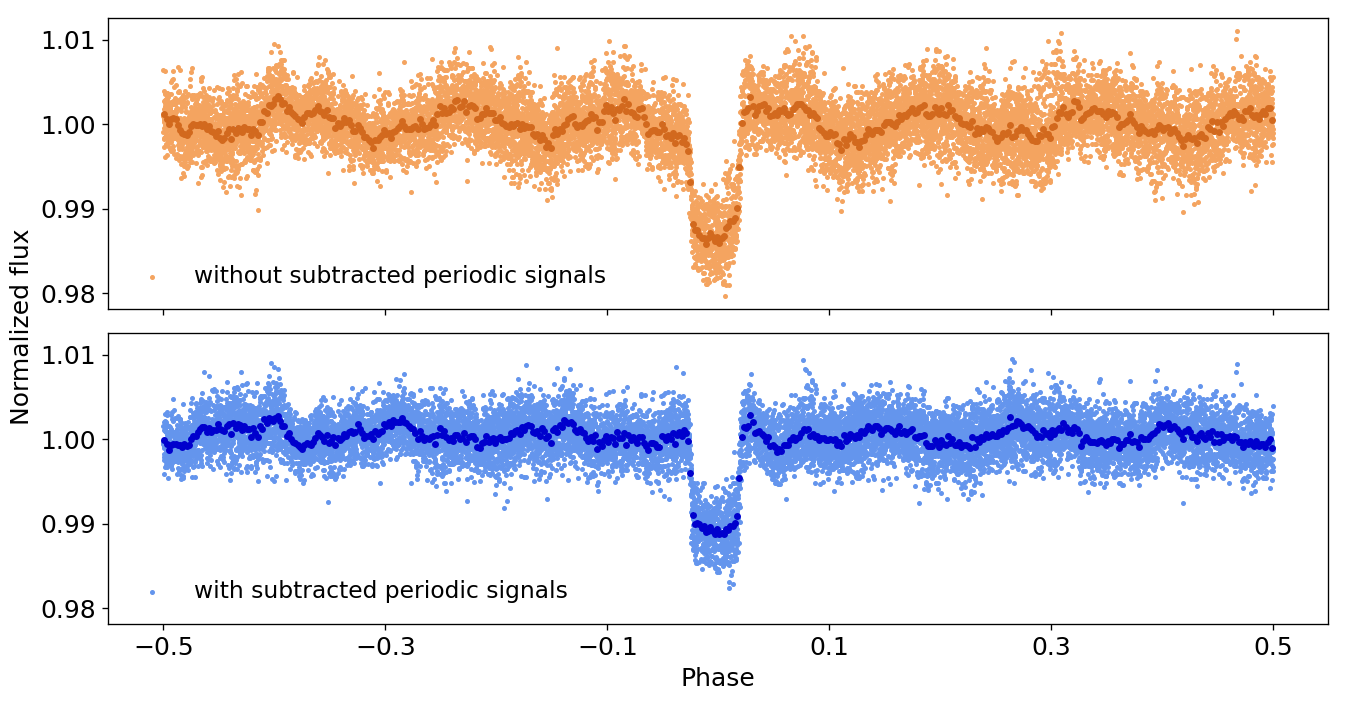}
    \caption{Phase-folded and binned TESS light curve of BD+37~4838 before (top) and after (bottom) subtraction of the selected periodic components identified with \texttt{Period04}.}
    \label{fig:toi-1414_before-after_freq}
\end{figure*}

After the experience with HD~210058, we ran the sampler on each sector individually. However, unlike in HD~210058, we did not perceive any visible or numerical evidence of a continual change in transit geometry (Fig.~\ref{fig:toi-1414_phasezoom}). Therefore, Table~\ref{tab:toi-1414_photo_params} lists only a single distribution for each parameter, its mean being the average of three separate values from three fitted sectors (Fig.~\ref{fig:toi-1414_triple_corner}), except for the BJD, which was taken from sector 16. Moreover, significant photometric variability due to the pulsations and rotation pattern (Fig.~\ref{fig:toi-1414_before-after_freq}, Sect.~\ref{section:pulsations}) prevented us from obtaining a reliable fit of the secondary transit and hence of the eccentricity and argument of periastron, which, as we observe, are strongly prior-dominated. We attempted to model the stellar variability using Gaussian process (GP) regression with a Matérn-3/2 kernel and a damped harmonic oscillator kernel implemented inside \texttt{allesfitter}. However, the GP models suppressed most of the photometric variability along with any possible secondary-eclipse signal, which therefore could not be reliably recovered, as well as the information on the eccentricity and argument of periastron. In the end, we were forced to assume a circular orbit.
\par
In an alternative approach, we corrected the light curve for any significant frequencies beforehand. We manually masked out transits and using \texttt{Period04} \citep{Period04doc} applied Fourier decomposition on our data, subsequently subtracting dominant periodic signals. However, even after identifying more than 20 frequencies in the residuals, their successive subtraction did not lead to a fully flattened light curve, as flux variations remain present (Fig.~\ref{fig:toi-1414_before-after_freq}). This indicates that the variability is not fully described by this sinusoidal model, for which we did not achieve sufficient improvement in the detectability of the secondary eclipse. Consequently, we conclude that the occultation is likely obscured by stellar variability and remains undetectable in the available photometry.
\par
During the initial step of the Doppler shadow modeling (Sect.~\ref{section:toi-1414_DT}), we noticed a visible offset between the predicted position of the Doppler shadow and the observed residual CCF maps. Since such an offset may be caused by an inaccurate transit ephemeris, we examined the available transit mid-times using an O-C analysis. The initially adopted period $P=3.019575\,\rm d$ produced a linear trend in the O-C diagram, indicating that the reference period was slightly underestimated. We therefore refitted the linear ephemeris and obtained an updated period of $P=3.0196527\,\rm d$. With this revised ephemeris, small residuals remain in the O-C diagram. These are, however, of the order of the timing uncertainties of the individual transit mid-times. Moreover, with only three points, this result is hardly statistically significant. For this, we decided to use \texttt{allesfitter} to compute a rapid fit of possible transit timing variations with a single reference period for all three sectors. This ephemeris analysis shows that the O-C scatter lies around zero, with most measurements consistent with zero within their uncertainties. Therefore, we cannot report any visible coherent nonlinear or periodic structure.
\par
As a sanity check, we repeated the photometric fitting with the period fixed at $P=3.0196527\,\rm d$. The results are given in the second posterior column of Table~\ref{tab:toi-1414_photo_params}.

\begin{table}[]
    \caption{Transit parameters of BD+37~4838~b.}
    \centering
    \begin{tabular}{lcc}
        \hline\hline
        \noalign{\smallskip}
        Parameter & Value & Value\\
        & Free $P$ & Fixed $P$ \\
        \noalign{\smallskip}
        \hline
        \noalign{\smallskip}
        No. transits & 22 & 22 \\
        $T_0$ (BJD-2457000) & $1751.41086_{-0.00027}^{+0.00035}$ & $1751.41092_{-0.00022}^{+0.00024}$ \\
        $P$ (d) & $3.019525_{-0.000036}^{+0.000021}$ & 3.0196527 \\
        $a/R_\star$ & $6.972_{-0.150}^{+0.123}$ & $6.923_{-0.144}^{+0.103}$ \\
        $R/R_\star$ & $0.0957_{-0.0012}^{+0.0011}$ & $0.0955_{-0.0036}^{+0.0013}$ \\
        $R$ (R\textsubscript{Jup}) & $1.473\pm0.067$ & $1.471\pm0.067$ \\
        $a$ (au) & $0.0512\pm0.0024$ & $0.0509\pm0.0024$ \\
        $i$ (deg) & $88.57_{-0.78}^{+0.56}$ & $88.14_{-0.72}^{+1.09}$ \\
        $b$ & $0.203_{-0.066}^{+0.088}$ & $0.223_{-0.127}^{+0.079}$ \\
        $T_{\mathrm{total}}$ (h) & $3.582_{-0.032}^{+0.030}$ & $3.582_{-0.019}^{+0.025}$ \\
        $T_{\mathrm{full}}$ (h) & $2.926_{-0.031}^{+0.030}$ & $2.924_{-0.028}^{+0.024}$ \\
        $q_{\mathrm{1}}$ & $0.121_{-0.044}^{+0.035}$ & $0.118_{-0.061}^{+0.078}$ \\
        $q_{\mathrm{2}}$ & $0.256_{-0.066}^{+0.045}$ & $0.266_{-0.075}^{+0.084}$ \\
        \noalign{\smallskip}
        \hline
    \end{tabular}
    \label{tab:toi-1414_photo_params}
\end{table}

\subsubsection{Pulsations of BD+37~4838} \label{section:pulsations}
\begin{figure*}
    \centering
    \includegraphics[width=0.9\linewidth]{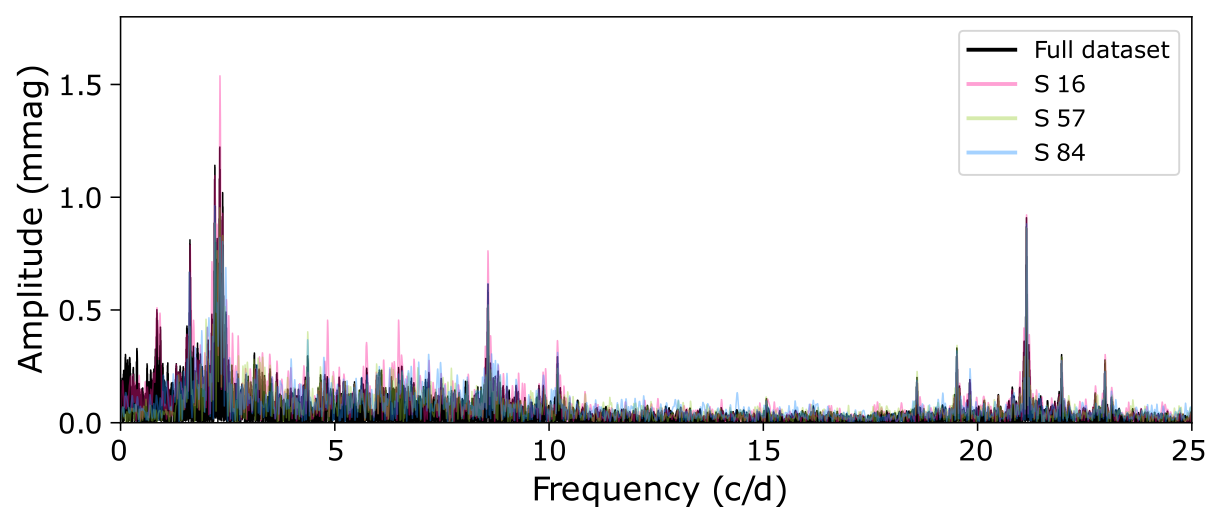}
    \caption{Frequency spectra based on a complete \texttt{SPOC} dataset (black line) and single sectors (color-coded).}
    \label{fig:TOI1414Frequency}
\end{figure*}

To investigate the pulsational properties of the system, we analyzed the TESS PDCSAP light curves produced by the \texttt{SPOC} pipeline \citep{Jenkins16} obtained with a 120 s cadence in Sectors 16, 57, and 84. All transit events were manually removed prior to analysis to avoid introducing spurious orbital frequencies. The combined dataset spans approximately 1870 days, corresponding to a Rayleigh frequency resolution of $5.35\times10^{-4}\,\rm c\,d^{-1}$. Frequencies were extracted using an iterative pre-whitening procedure implemented in \texttt{Period04}. The analysis was performed both on the combined dataset and on the individual TESS sectors to evaluate the long-term stability of the pulsation modes. In total, 27 statistically significant frequencies with S/N>4 \citep{Breger1993} were identified (Table~\ref{tab:TOI-1414Frequnencies}). The frequency spectrum exhibits both low-frequency $\gamma$~Dor pulsations (predominantly around $2.3\,\rm c\,d^{-1}$) and higher-frequency $\delta$~Sct modes between approximately 8 and $26\,\rm c\,d^{-1}$ (Fig.~\ref{fig:TOI1414Frequency}), indicating that the primary is a hybrid $\gamma$~Dor-$\delta$~Sct pulsator.

The dominant low-frequency mode, $f_{1}=2.3159\,\rm c\,d^{-1}$, is located close to the seventh harmonic of the orbital frequency ($f_{1}=6.993\,f_{\rm orb}$), producing the characteristic beating pattern visible in the upper light curve in Fig.~\ref{fig:toi-1414_before-after_freq}. At first glance, this near-commensurability suggests a possible tidal origin, since tidally excited oscillations are expected to occur at integer multiples of the orbital frequency and remain phase-locked to the orbital motion \citep{Fuller2017}. Moreover, the stellar parameters ($T_{\rm eff}=7128\,\rm K$, $M=1.59~\mathrm{M_{\odot}}$, $R=1.588\,\mathrm{R_{\odot}}$) place the primary inside the $\gamma$~Dor instability strip, while its rapid rotation ($v\sin i_\star=81\,\rm km\,s^{-1}$) makes a tidally perturbed prograde sectoral quadrupole mode ($l=2$, $m=+2$) a plausible initial interpretation \citep{Saio2018,Li2020}.
\par
Nevertheless, several observational facts argue against a classical tidally excited oscillation. The measured frequency differs from the exact seventh orbital harmonic ($7f_{\rm orb}=2.31825\,\rm c\,d^{-1}$) by $0.00235\,\rm c\,d^{-1}$, which is more than four times the Rayleigh frequency resolution. Furthermore, the frequency detected near $2.31\,\rm c\,d^{-1}$ varies between individual TESS sectors ($2.3111-2.3176\,\rm c\,d^{-1}$), whereas several neighboring pulsation modes remain remarkably stable. This indicates that the observed peak most likely represents a group of closely spaced modes with variable amplitudes rather than a single coherent orbitally locked oscillation. In addition, no other detected frequency lies close to an integer multiple of the orbital frequency, which argues against a broader pattern of tidal harmonics. In contrast, the low-frequency modes form a sequence near 2.20, 2.25, 2.32, 2.39, and $2.46\,\rm c\,d^{-1}$ with increasing frequency spacing, a characteristic signature of prograde gravito-inertial modes in rapidly rotating $\gamma$~Dor stars \citep{Saio2018}. Although a detailed period-spacing analysis is beyond the scope of this work, the available evidence favors an interpretation in which the dominant low-frequency variability arises from intrinsic rotationally modified $\gamma$~Dor g-modes, while the higher-frequency peaks correspond to $\delta$~Sct p modes. Consequently, the proximity of $f_{1}$ to the seventh orbital harmonic is most likely coincidental, although a weak tidal influence on mode selection or visibility cannot yet be completely excluded.

\subsubsection{Doppler tomography} \label{section:toi-1414_DT}
Similarly to HD~210058, we had no prior knowledge of this system's spectroscopic parameters, for which we kept the same wide priors, except for the systemic velocity $\gamma$, $\mathcal{N}(-1.39, 6.90)\,\rm km\,s^{-1}$, again estimated using \texttt{iSpec}. This time, however, we did not need to adjust the uncertainty, since it did not deviate significantly from the mean value, while still showing agreement between the real planetary trace and the one modeled using results from the light curve analysis. 
\par
Using the updated period, $P=3.0196527\,\rm d$, the tomographic signal appears broadly consistent with a prograde trajectory across the stellar line profile. The best fitting model (Table~\ref{tab:DSs_priors_posteriors}) favors a projected obliquity of $\lambda = 56.9^{+1.4}_{-2.3}\,\rm deg$, suggesting that BD+37~4838~b may also be on a misaligned orbit. Such geometry could (as in the case of HD~210058 and assuming the same $k_2$ and an equator-on view of the star) be consistent with the analytical estimate $\dot{i_\mathrm{orb}}=0.085\,\rm deg\,yr^{-1}$. This value is, however, smaller than the uncertainty of the fitted slope of the three-sector $i_\mathrm{orb}$ measurements, $\sigma=0.12\,\rm deg\,yr^{-1}$, which is too weak of a constraint to provide a detection.

\begin{figure*}
    \centering
    \includegraphics[width=\linewidth]{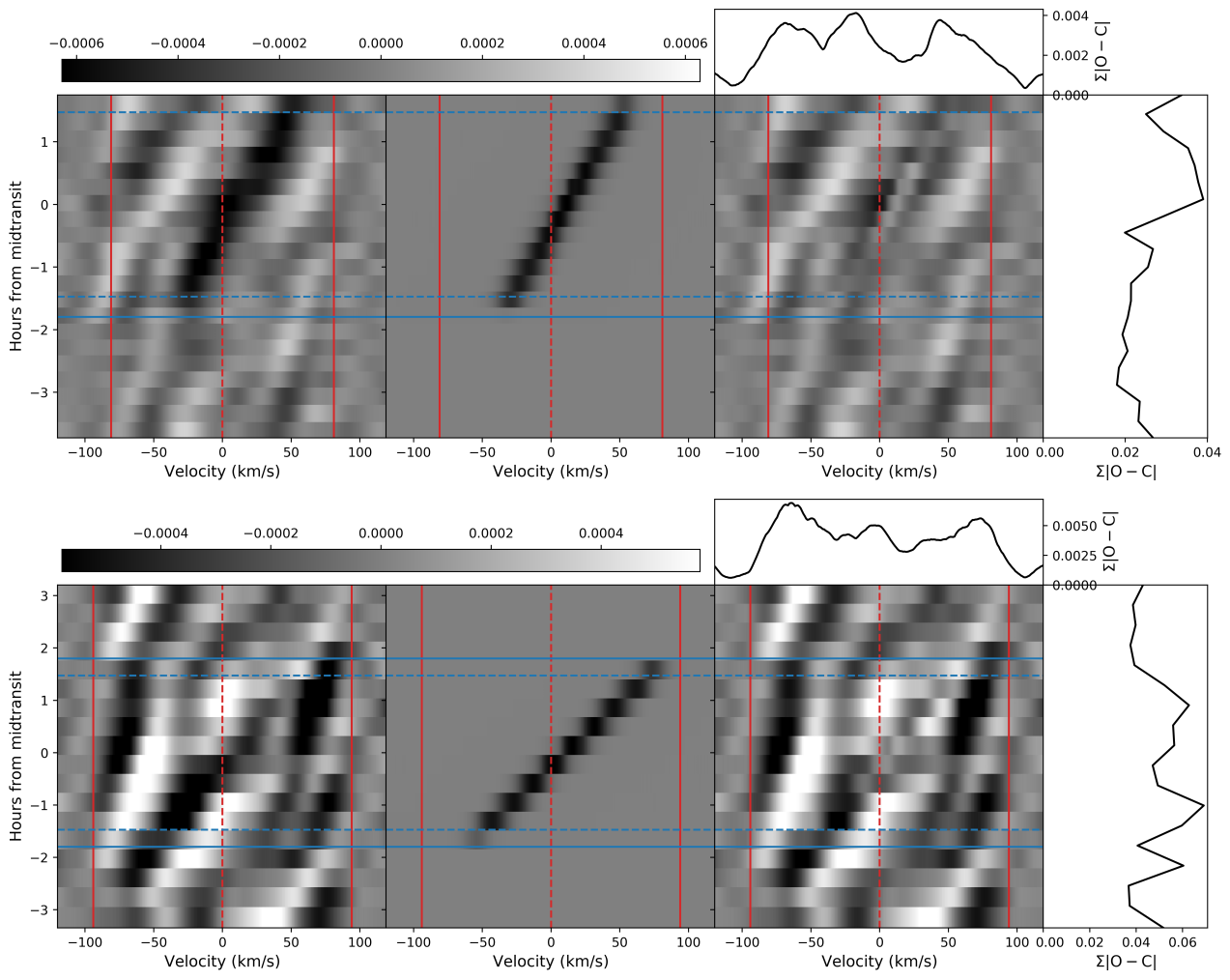}
    \caption{Doppler tomography of BD+37~4838. \textit{Top}: HARPS-N. \textit{Bottom}: PEPSI. Each of the two subfigures contains observed line-profile residuals (left), the best fitting Doppler-shadow model (middle), and the residuals after model subtraction (right). Top and far-right panels: Sums of the absolute residuals along the time and velocity axes, respectively. The vertical red lines mark the approximate limits of the stellar rotational profile, while the dashed blue lines indicate the extent of the transit. The significant residual structure remaining after model subtraction is consistent with stellar variability.}
    \label{fig:toi-1414_DT}
\end{figure*}

Subsequently, we examined the PEPSI observations. The spectra were already normalized, while the CCFs were extracted following the same procedure as for the HARPS-N observations. Applying the same \texttt{tracit} tomographic analysis yields a somewhat higher projected stellar rotational velocity of $94\,\rm km\,s^{-1}$ (compared to $81\,\rm km\,s^{-1}$ for the HARPS-N data) and a lower projected obliquity $\lambda\approx 43.6\,\rm deg$. However, the expected Doppler shadow feature in the PEPSI data (Fig.~\ref{fig:toi-1414_DT}, lower panel) is considerably less distinct relative to the pulsational line-profile variability than the Doppler shadow detected in the HARPS-N data (Fig.~\ref{fig:toi-1414_DT}, upper panel). Given the comparable amplitudes and partially overlapping trajectories of the pulsational structures, the inferred PEPSI signal may be biased, or partly produced, by stellar variability, which would likely enhance the degeneracy between the trajectory, width, and projected rotational velocity. Therefore, although the PEPSI solution is broadly compatible with the prograde, misaligned geometry, we prefer not to treat it as an independent quantitative measurement of the projected obliquity.
\par
Both residual maps show that substantial line-profile variability remains after subtracting the transit model. Since the pulsation pattern overlaps with the expected velocity range of the planetary Doppler shadow, it is uncertain whether, and to what extent, the inferred obliquity is affected by it. As a diagnostic test, we attempted to isolate the Doppler shadow from the pulsations using the Fourier filtering methodology introduced by \cite{Johnson2015}. In the case of BD+37~4838, both the Doppler shadow and the pulsations are prograde, meaning that they move in the same direction across the line profile. This is in contrast to previous cases such as WASP-33~b \citep{Johnson2015} or WASP-167~b (KELT-13~b) \citep{Temple2017}, where the planetary orbit is retrograde while the pulsations are prograde, leading to a clean separation in frequency space between these two components. In BD+37~4838, the Doppler shadow and pulsations occupy the same regions of frequency space, and we were unable to construct a mask to cleanly separate these components.

\subsection{Mass estimation}
We discuss the possibility that these candidate exoplanetary companions may, in reality, be brown dwarfs rather than hot Jupiters. Brown dwarfs are substantially more massive than giant planets, but their radii are not proportionally larger. In fact, they are broadly Jupiter-sized, or even smaller \citep{Burrows2001}. A useful example is HIP~33609~b, an eccentric transiting brown dwarf with a mass of $\sim68\,\mathrm{M_{Jup}}$ but a radius of only $1.58\,\mathrm{R_{Jup}}$ \citep{Vowell_2023}. This behavior arises because substellar objects contract with time and their internal structure is increasingly affected by electron degeneracy. At a fixed mass, the radius generally decreases with age. However, at later evolutionary stages, objects with very different masses can have similar radii, and the mass-radius relation may even become non-monotonic. According to \citet{Burrows2001}, for older substellar objects in the broad mass range of $0.3-70\,\mathrm{M_{Jup}}$, radii are nearly independent of mass to within about $30\,\%$. This near-consistency of the radius is closely related to the radius inversion discussed by \citet{Sorahana_2013}, who note that more massive brown dwarfs, although larger when young, can become smaller than less massive objects beyond $\sim10^8\,\rm yr$.
\par
Therefore, the measured or inferred radius alone cannot determine the correct classification of such an object. We thus turn to relevant mass estimates to evaluate whether our candidates align better with hot Jupiters or with objects in the brown-dwarf mass regime.

\subsubsection{Directly inferred RV semi-amplitude}
A widely used formula that connects the companion mass ($M_\mathrm{b}$) with other observable quantities is the equation for the RV semi-amplitude:
\begin{equation}
    K = \left(\frac{2\pi G}{P}\right)^{1/3}\frac{M_\mathrm{b}\sin i}{\left(M_\star + M_\mathrm{b}\right)^{2/3}}\frac{1}{\sqrt{1-e^2}}.
\end{equation}
Unfortunately, this is very hard to determine for rapidly rotating stars since the RV uncertainty rapidly increases with increasing $v\sin{i_\star}$. Moreover, our spectroscopic observations do not cover the whole orbit, only a short phase around transit. Despite this, we inferred the $K$ of both HD~210058 and BD+37~4838 directly from the tomography map and observe the following behavior: when left free, the posterior is non-convergent, and it drifts toward very low values. When a Gaussian prior is set, the posterior remains tightly prior-dominated. Both cases produce basically identical posteriors for the other parameters, except for $v\sin i_\star$ in HD~210058, for which higher prior semi-amplitudes produce a higher additional posterior peak around $\sim150\,\rm km\,s^{-1}$. However, a fast spectral fit in \texttt{iSpec} confirms this value to be unrealistically elevated. This indicates that the transit CCF dataset primarily constrains the line-profile distortion geometry rather than the orbital stellar semi-amplitude. We further confirmed this independence by running the sampler without any prior on $K$. Again, the resulting posteriors show no sensitivity to such a choice.
\par
\begin{figure}
    \centering
    \includegraphics[width=\columnwidth]{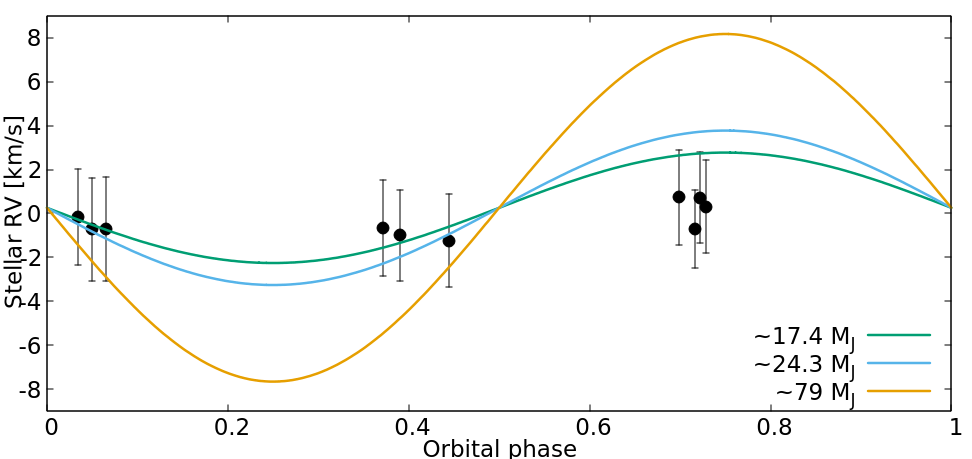}
    \caption{RV of BD+37 4838 as a function of orbital phase. The scattered black points represent a small set of RVs extracted from the NEID spectra. The green and blue curves correspond, in this order, to the stellar RVs induced by the fitted 1$\sigma$ and 2$\sigma$ upper boundary companion masses on a circular orbit. The orange curve represents a low-mass M-type star.}
    \label{fig:toi1414-rvs}
\end{figure}

 
As a next step, we derived RVs from the available NEID spectra (Fig.~\ref{fig:toi1414-rvs}), for which we employed a custom iSpec-based cross-correlation procedure broadly analogous to that used for the extraction of the CCFs in the Doppler tomographic analysis. In this case, however, the observed spectra were cross-correlated against a synthetic template rotationally broadened to approximately match the line widths of the target star, thereby providing a closer representation of the observed spectral morphology. We tested two approaches to cross-correlation: one over a broad, combined wavelength range, and another in which the spectrum was divided into several smaller wavelength intervals, with the resulting velocities subsequently combined. Unfortunately, neither of the two cross-correlation procedures achieves substantially higher precision. Due to the broad and variable line profiles associated with the rapidly rotating and pulsating host star, the uncertainties of the final RV measurements remain at $\sim2\,\rm km\,s^{-1}$.
\par
Using \texttt{allesfitter}, we attempted to constrain the RV semi-amplitude, $K$, while fixing the orbital ephemeris to the photometric solution. The resulting posterior distribution is highly non-Gaussian and decreases from its maximum at the lower prior boundary, $K=0$. Although the central $68\,\%$ posterior interval spans $0.43-2.51\,\rm km\,s^{-1}$, its positive lower bound is induced by the restriction $K\geq0$ and must not be interpreted as a detection of a nonzero RV amplitude. We therefore report only a one-sided $95\,\%$ credible upper limit of $K<3.51\,\rm km\,s^{-1}$ and, similarly, an upper mass limit of $M_{\rm b}<24.34\,\mathrm{M_{Jup}}$. Although the poor phase coverage and large uncertainties preclude any meaningful estimation of the orbital eccentricity, leaving it essentially unconstrained by both the photometric and RV data, our result at least confirms the companion's substellar nature. Considering the low-mass boundary of an M dwarf, $0.075\,{\rm M_{\odot}}$ \citep{Henry24}, we would expect the semi-amplitude to reach as high as $\sim7.9\,\rm km\,s^{-1}$ (Fig.~\ref{fig:toi1414-rvs}).

\subsubsection{Ellipsoidal variations}
One effect shaping the overall photometric light curve may be due to ellipsoidal variations. These arise from the tidal deformation of a star by a close companion, producing photometric variability as the projected stellar surface changes during orbit. Assuming that $\left(\frac{R_\star}{a}\right)$ is small, the amplitude of this effect is given by \citep{Mazeh10} as
\begin{equation}
    A_{\mathrm{ellip}} \approx \alpha_{\mathrm{ellip}} \frac{M_b}{M_\star} \left(\frac{R_\star}{a}\right)^3,
\end{equation}
where 
\begin{equation}
    \alpha_{\mathrm{ellip}} = 0.15 \frac{(15 + u)(1 + g)}{3 - u}.
\end{equation}
Here, $g$ is the stellar gravity-darkening coefficient, and $u$ is the linear limb-darkening coefficient. Since the ellipsoidal modulation is slightly visible in the phase curve of HD~210058, we applied this approximation to estimate the mass of its companion. From the phase-curve fit, we obtain $A_{\mathrm{ellip}}=(0.0692\pm0.0098)\times10^{-3}$ and, in addition to the quadratic limb-darkening coefficients used in the light-curve model, we estimate the corresponding linear coefficient to be $u=0.420\pm0.020$. We did not obtain a reliable fit of the gravity-darkening coefficient itself and therefore adopted $g=1$, which corresponds to the classical von Zeipel gravity-darkening law for stars with radiative envelopes \citep{VZeipel1924}. To propagate the uncertainties in the derived companion mass, we performed a Monte Carlo calculation with $5\times10^5$ draws, in which all input parameters were sampled from normal distributions centered on their measured values with standard deviations corresponding to their $1\sigma$ uncertainties. Using this procedure, we obtain a companion mass $M_{\rm b} = 10.0^{+2.5}_{-2.2}\,\mathrm{M_{Jup}}$, which favors a massive hot-Jupiter interpretation rather than a brown dwarf. Using this mass, we additionally estimate the RV semi-amplitude to be $K\approx892.23\,\rm m\,s^{-1}$.

\subsubsection{Doppler beaming}
This photometric effect is caused by the radial motion of a star, which relativistically enhances its observed flux as it moves toward the observer and reduces it as it recedes. As described in \cite{Barbier21}, in simplified form,
\begin{equation}
    A_{\rm beam} = (3-\alpha_{\rm beam})\frac{K}{c},
    \label{eq:beaming}
\end{equation}
where
\begin{equation}
    \alpha_{\rm beam} = \frac{e^x(3-x)-3}{e^x-1},
\end{equation}
with 
\begin{equation}
    x = \frac{hc}{\lambda k T_{\rm eff}}.
\end{equation}
Here, $\lambda$ is taken as the TESS central wavelength of $786.5\, \rm nm$. Using the same fitting setup for HD~210058 as in the previous case, we retrieve a Doppler beaming amplitude $A_{\mathrm{beam}}=(0.0240\pm0.0065)\times10^{-3}$. Next, we estimated the companion mass using the same Monte Carlo approach as described above, with the same seed from the random number generator, and again performed $5\times10^5$ draws. From Eq. (\ref{eq:beaming}), we infer an RV semi-amplitude $K = 3152.9^{+867.9}_{-853.1}\,{\rm m\,s^{-1}}$, which corresponds to a companion mass computed from the standard RV semi-amplitude equation, $M_{\rm b} \approx 34.6^{+10.7}_{-9.9}\,\mathrm{M_{Jup}}$. This is obviously strongly inconsistent with the previous mass estimate of $\sim10\,\mathrm{M_{Jup}}$. One possible caveat is that Eq. (\ref{eq:beaming}) assumes a sinusoidal RV variation; therefore, since the Doppler-boosting signal follows the instantaneous stellar RV, it represents a simplified circular-orbit approximation. HD~210058~b is eccentric, which means that this simplification could have a measurable effect; however, this alone is unlikely to explain a factor exceeding three in the inferred companion mass.
\par
A similar caution about interpreting the Doppler beaming amplitudes was raised by \cite{Barbier21}, who compared theoretical, fitted, and published beaming amplitudes and RV values for several Kepler systems, finding significant discrepancies in most cases. Although their comparison is not directly equivalent to the internal comparison between the beaming and ellipsoidal amplitudes performed here, they note that Doppler beaming can be sensitive to the adopted model, harmonics, or instrumental precision. 
\par
As an additional consistency test, we fixed the Doppler beaming amplitude to the value predicted for a companion with its mass inferred from the ellipsoidal variations, that is, $A_{\mathrm{beam}}=0.00685\times10^{-3}$. Despite this modification, the recovered ellipsoidal amplitude remains essentially unchanged. This indicates that the mass estimate based on the ellipsoidal variations is largely insensitive to the assumed Doppler beaming amplitude. In contrast, the latter appears to be much more model-dependent, making it a less reliable mass indicator, at least in the present case.

\section{Discussion}
The measured sky-projected obliquities of HD~210058~b and BD+37~4838~b provide useful clues to their possible dynamical histories, although the interpretation is not straightforward. Recently, \citet{Wang26b} found that the distribution of systems in the eccentricity-obliquity plane depends strongly on planet mass. In their single-star sample, sub-Saturns can be both eccentric and misaligned. Jupiter-mass planets are misaligned preferentially on nearly circular orbits, while super-Jupiters and brown dwarfs appear largely aligned over the full eccentricity range. In this context, HD~210058~b is close to, but not necessarily in strong tension with, the high-mass population discussed by \citet{Wang26b}, since the projected obliquity remains below their adopted threshold of $\lambda=40\,\rm deg$ for strongly misaligned systems. A complementary perspective is offered by \citet{Wang26c}, who predict that super-Jupiters and brown dwarfs orbiting hot stars may exhibit significant misalignments at sufficiently small orbital separations ($a_{\mathrm{final}}/R_{\mathrm{p}}\lesssim100$), where $a_{\mathrm{final}}=a(1-e^2)$ represents the expected semi-major axis after tidal circularization, assuming approximate conservation of orbital angular momentum. With $a_{\mathrm{final}}/R_{\mathrm{p}}\approx57$, HD~210058~b falls within this regime and supports this scenario. The combination of substantial eccentricity and nonzero projected obliquity nevertheless points to a dynamically nontrivial history, especially in the absence of a detected planetary or stellar companion.
\par
Possible explanations of orbital excitation are planet-planet scattering, secular interactions, or perturbation from an additional undetected companion \citep[e.g.][]{Fabrycky2007, Giacalone2017}. Although the latter could provide a plausible explanation for maintaining the eccentricity of a planet as massive as HD~210058~b, it is disfavored by the star's RUWE value, which does not indicate any significant astrometric deviations from a single-star solution -- at least not on the current DR3 data time baseline. Moreover, we must keep in mind the companion's measured projected obliquity of $-32\,\rm deg$. If this spin–orbit misalignment was not produced by post-formation dynamical interactions, it could instead reflect a primordial misalignment of the system. However, a primordial origin of the obliquity alone would not naturally explain the planet's significant orbital eccentricity \citep{Batygin2013}, leaving room for an additional excitation mechanism.
\par
All the listed mechanisms for orbital excitation are often associated with high-eccentricity migration, in which the orbit is first driven to high eccentricity and subsequently circularized by tides. The present-day eccentricity of HD~210058~b suggests that such tidal evolution may yet be incomplete. Using our system parameters, we estimate the timescale of its orbital circularization using 
\begin{equation}
    \tau_{circ}\sim e/(\mathrm{d}e/\mathrm{d}t),
\end{equation}
where
\begin{equation}
    \frac{\mathrm{d}e}{\mathrm{d}t}=\frac{\mathrm{d}E}{\mathrm{d}t}\frac{a(1-e^2)}{GM_\star M_be}.
\end{equation}
Here, $E$ is the orbital energy of the planet, which is further elaborated on in \cite{Rice2022}. Taking into account the expected tidal dissipation efficiency, which spans two orders of magnitude \citep{Penev2018}, the circularization timescale ranges from $\sim10\,\mathrm{Myr} - 1\,\mathrm{Gyr}$, according to our calculations. This implies that the system is either relatively young (or at least that the planet is dynamically young) or that an additional mechanism acts to maintain the observed eccentricity. Therefore, we measured the age of HD~210058 by searching for co-moving stars using the \texttt{FriendFinder}\footnote{\url{https://github.com/adamkraus/comove}} tool \citep{THYMEV}. This tool leverages \textit{Gaia} DR3 data to identify co-moving stellar populations: it projects the 6D coordinates of the target onto a 5D space, then predicts every nearby star's proper motions and positions by assuming a match to our target star. The predicted tangential velocity $v_{\rm{tan}}$ and RV can be compared directly to \textit{Gaia} outputs. For our search, we adopted generous criteria of $v_{\rm{tan}}<4\,\rm km\,s^{-1}$ and a 3D distance $<40\,\rm pc$, yielding 449 candidate co-moving stars.
\par 
The resulting population is sparse and heavily contaminated by nearby stars that are co-moving by chance. However, there are four reasons to think the group is real: (1) there is an over-density of stars with radial velocities matching the predicted value (Fig.~\ref{fig:drv}); (2) $48\,\%$ of candidate co-moving stars with $G<14$~mag have periods in the TARS catalog \citep{Boyle2026}, compared to only 20--25$\,\%$ for random draws of nearby stars with no kinematic association; (3) rotations with good RV agreement and temperatures suitable for gyrochronology yield a consistent age of about 200--300~Myr (Fig.~\ref{fig:prot}); and (4) the M dwarfs preferentially lie above the main sequence in the color-magnitude diagram, which is consistent with a young age and rare in random field draws. Unfortunately, these points are more suggestive than conclusive (random field draws such as this are rare, not impossible), so we consider this group likely. However, confirmation would require additional data, such as lithium abundance \citep{Soderblom2013}, more RVs, or \textit{Gaia} DR4 data.
\par
Assuming the group is real, we estimated its age using a combination of \texttt{gyro-interp} \citep{Bouma2023} on the rotation periods, an excess variability-based age method on the \textit{Gaia} photometry \citep[which uses variability as an age metric,][]{Barber2023}, and an isochrone fit with a mixture model to account for contamination \citep{THYMEVI}. More details on this method and how the multiple age metrics are merged can be found in \citet{Barber2026}. The final age determination is $181\pm35\,\rm Myr$, with each individual estimate consistent within $<1.5\sigma$. Compared with the circularization timescale of the planetary orbit, this favors a scenario in which the process may still be ongoing.
\par
The implications of the observed obliquity for the dynamical evolution of BD+37~4838~b are more difficult to assess, because without a reliable mass measurement it cannot yet be placed in the mass-dependent $e$-$\lambda$ framework. If BD+37~4838~b is a Jupiter-mass planet on a nearly circular orbit, its large projected obliquity would be broadly consistent with the population of misaligned hot Jupiters. If, however, it is a super-Jupiter or brown dwarf, its obliquity would make it more unusual compared with the predominantly aligned high-mass population in single-star systems. In this latter case, an undetected stellar companion or a primordial misalignment mechanism could provide a natural explanation.

\section{Summary}
In this work, we presented a photometric and spectroscopic analysis of the HD~210058 (TOI-1355) and BD+37~4838 (TOI-1414) systems. We find that HD~210058~b lies on an eccentric orbit around its young ($\sim180\,\rm Myr$) host star, while exhibiting detectable nodal precession with an estimated precession period of approximately 860 years. Although this period is far longer than the current observational baseline, changes in the transit geometry are measurable over the five years of archival TESS data and will remain observable for only about six more years, after which transits are expected to cease. This makes HD~210058 a very timely target for some types of follow-up observations. For example, \cite{Hord24} previously identified this star's companion as a potentially favorable target for both transmission and emission spectroscopy. However, the new mass value presented in this paper implies a high surface gravity and, consequently, a relatively small atmospheric scale height, hindering the detection of transmission features. Thermal-emission spectroscopy may therefore offer a more promising route for further investigation.
\par
Next, we identified BD+37~4838 as a hybrid $\gamma$~Dor-$\delta$~Sct pulsator, with its dominant pulsation frequency located close to the seventh harmonic of the orbital frequency. Although this proximity could indicate a weak tidal influence on the stellar oscillations, we consider coincidence to be the more likely explanation. Because the stellar pulsations dominate the phase curve of BD+37~4838 and obscure the secondary eclipse, we are unable to reliably recover either ellipsoidal-variation or the Doppler-beaming signal and thus cannot derive a photometric mass estimate of the orbiting companion. A series of spectroscopic observations for this system was also analyzed, although the large uncertainties on the resulting RV measurements precluded detection of the orbital signal, yielding only an upper mass limit of $24.34\,\mathrm{M_{Jup}}$. The situation was more favorable for HD~210058~b, for which ellipsoidal variation provided the most reliable mass constraint among the phase-curve components. From this signal, we estimated the mass of HD~210058~b to be $M_{\rm b} = 10.0^{+2.5}_{-2.2}\,\mathrm{M_{Jup}}$, placing it among the most massive hot Jupiters, just below the conventional brown-dwarf boundary.
\par
Our Doppler tomography analysis yields sky-projected spin-orbit angles of $\lambda=-32_{-6}^{+4}\,\rm deg$ for HD~210058~b and $\lambda=56.9^{+1.4}_{-2.3}\,\rm deg$ for BD+37~4838~b, indicating that both planets reside on misaligned orbits. Interestingly, both host stars exhibit similar projected rotational velocities, $v\sin i_\star \approx 81\,\rm km\,s^{-1}$, although the value inferred for BD+37~4838 may be slightly affected by pulsation-induced line-profile variability. For neither system was it possible to determine the stellar rotation period ($P_\mathrm{rot}$) and, hence, the stellar inclination ($i_\star$). Consequently, the true 3D spin-orbit angle ($\psi$) remains unconstrained (Eqs.~\ref{eq_psi} and \ref{eq_prot}).

\begin{acknowledgements}

We would like to thank the anonymous referee for their kind review, as well as the editor for pointing out several stylistic imperfections. ZB and PK acknowledge the support of a GACR grant 22-30516K; JZ and MS acknowledge the support of the PRODEX agreement no. 400012791; the research of PG was supported by the Slovak Research and Development Agency under contract No. APVV-24-0160. We thank the KELT and PEPSI teams for providing the data on BD+37~4838. We also acknowledge the use of TESS High Level Science Products (HLSP) produced by the Quick-Look Pipeline (QLP) at the TESS Science Office at MIT, which are publicly available from the Mikulski Archive for Space Telescopes (MAST). Funding for the TESS mission is provided by NASA's Science Mission directorate. Additionally, we acknowledge the use of the Exoplanet Follow-up Observation Program (ExoFOP) website, operated by the California Institute of Technology under contract with NASA as part of the Exoplanet Exploration Program. This research also used the facilities of the Italian Center for Astronomical Archive (IA2) operated by INAF at the Astronomical Observatory of Trieste. The programs that observed HD~210058 and BD+37~4838 can be found under IDs of ${\rm A44TAC\_38}$ and ${\rm OPT15B\_19}$, respectively. ZB also thanks E. Knudstrup for providing answers on several questions regarding the usage of \texttt{tracit}, and J. Lipták for a useful discussion about exoplanetary masses.
\end{acknowledgements}

\bibliographystyle{aa}
\bibliography{references}

\begin{appendix}

\onecolumn
\section{Photometry of BD+37~4838, posterior distribution}

\begin{figure*}[ht!]
    \centering
    \includegraphics[width=1\linewidth]{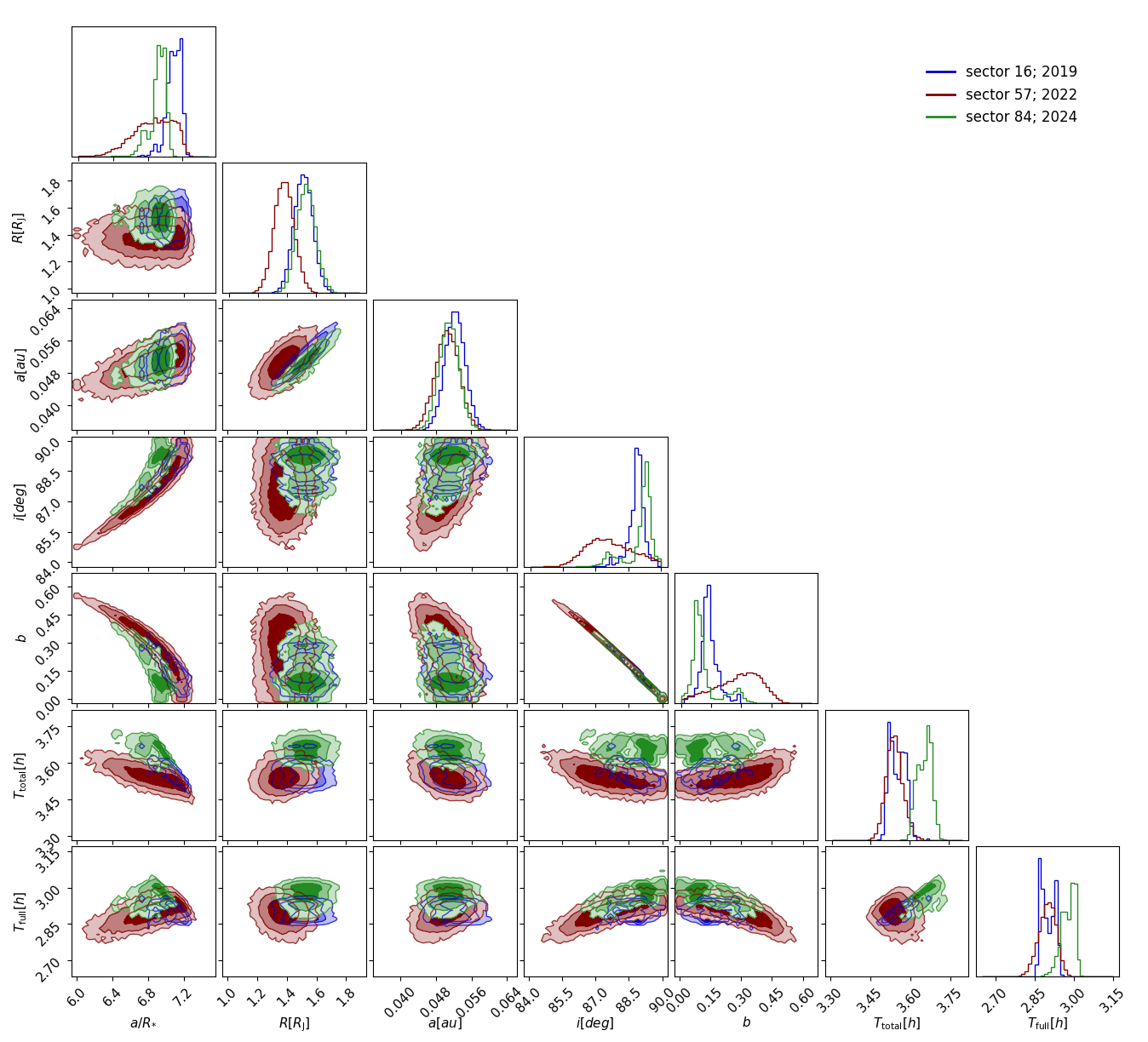}
    \caption{Corner plot of the posterior distributions of the fitted parameters obtained from the three independent BD+37~4838 TESS measurements. The contours correspond to the 1, 2 and $3\sigma$ credible intervals.}
    \label{fig:toi-1414_triple_corner}
\end{figure*}

\FloatBarrier 

\clearpage
\section{Precessing exoplanetary systems}

\begin{table*}[ht!]
    \centering
    \caption{Comparison of known precessing exoplanetary systems.}
    \resizebox{\textwidth}{!}{%
    \begin{tabular}{lccccc}
        \hline\hline
        \noalign{\smallskip}
        System
        & KOI-13
        & WASP-33
        & KELT-9
        & TOI-1518
        & HD~210058 \\
        
        \noalign{\smallskip} \hline \noalign{\smallskip}
        \multicolumn{6}{c}{\textit{Host star parameters}} \\
        \noalign{\smallskip} \hline \noalign{\smallskip}

        $T_{\mathrm{eff}}$ [K]
        & $8482.0^{+173.6}_{-197.1}$ \textsuperscript{(1)}
        & $7430 \pm 100$ \textsuperscript{(7)}
        & $9270^{+240}_{-180}$ \textsuperscript{(12)}
        & $7300 \pm 100$  \textsuperscript{(15)}
        & $9218^{+835}_{-437}$ \\

        $M_{\star}$ [${\rm M_{\odot}}$]
        & $1.670^{+0.074}_{-0.096}$ \textsuperscript{(1)}
        & $1.495 \pm 0.031$ \textsuperscript{(7)}
        & $2.32 \pm 0.16$ \textsuperscript{(12)}
        & $1.79 \pm 0.26$  \textsuperscript{(15)}
        & $2.34^{+0.53}_{-0.24}$ \\

        Age [Gyr]
        & $1.12^{+0.13}_{-0.07}$ \textsuperscript{(1)}
        & $<0.4$ \textsuperscript{(7)}
        & $0.45^{+0.14}_{-0.13}$ \textsuperscript{(12)}
        & ---
        & --- \\

        $v\sin i_{\star}$ [km\,s$^{-1}$]
        & $65 \pm 10$ \textsuperscript{(2)}
        & $90 \pm 10$ \textsuperscript{(7)}
        & $111.8 \pm 1.0$ \textsuperscript{(12)}
        & $85.1 \pm 6.3$  \textsuperscript{(15)}
        & $81^{+9}_{-11}$ \\
        
        \noalign{\smallskip} \hline \noalign{\smallskip}
        \multicolumn{6}{c}{\textit{Companion parameters}} \\
        \noalign{\smallskip} \hline \noalign{\smallskip}

        $M_{\mathrm{p}}$ [$\mathrm{M_{Jup}}$]
        & $9.28 \pm 0.16$ \textsuperscript{(3)}
        & $3.28 \pm 0.73$ \textsuperscript{(8)}
        & $2.88 \pm 0.84$ \textsuperscript{(13)}
        & $1.83 \pm 0.47$ \textsuperscript{(16)}
        & $9.95^{+2.49}_{-2.17}$ \\

        $R_{\mathrm{p}}$ [$\mathrm{R_{Jup}}$]
        & $1.52^{+0.19}_{-0.24}$ \textsuperscript{(1)}
        & $1.497 \pm 0.045$ \textsuperscript{(7)}
        & $1.891^{+0.061}_{-0.053}$ \textsuperscript{(13)}
        & $1.875 \pm 0.053$ \textsuperscript{(15)}
        & $1.425 \pm 0.055$ \\

        $P$ [d]
        & $\sim1.76359$ \textsuperscript{(1)}
        & $\sim1.21987$ \textsuperscript{(7)}
        & $\sim1.48112$ \textsuperscript{(13)}
        & $\sim1.90260$ \textsuperscript{(15)}
        & $\sim2.17023$ \\

        $a$ [au]
        & $\sim0.03641$ \textsuperscript{(3)}
        & $\sim0.02555$ \textsuperscript{(7)}
        & $\sim0.0346$ \textsuperscript{(13)}
        & $\sim0.0389$ \textsuperscript{(15)}
        & $\sim0.0408$ \\

        $a/R_{\star}$
        & $4.5007^{+0.0039}_{-0.0040}$ \textsuperscript{(3)}
        & $3.788^{+0.084}_{-0.080}$ \textsuperscript{(7)}
        & $3.153 \pm 0.011$ \textsuperscript{(13)}
        & $4.291^{+0.057}_{-0.061}$ \textsuperscript{(15)}
        & $4.740 \pm 0.091$ \\

        $e$
        & $\sim0$ \textsuperscript{(3)}
        & $0$ \textsuperscript{(8)}
        & $0$ \textsuperscript{(13)}
        & $<0.01$ \textsuperscript{(15)}
        & $0.2293^{+0.0052}_{-0.0044}$ \\

        $\lambda$ [deg]
        & $58.6 \pm 2$ \textsuperscript{(4)}
        & $-111.64 \pm 0.28$ \textsuperscript{(9)}
        & $-84.8 \pm 0.3$ \textsuperscript{(14)}
        & $240.34^{+0.93}_{-0.98}$ \textsuperscript{(15)}
        & $-32^{+4}_{-6}$ \\

        $\psi$ [deg]
        & $60 \pm 2$ \textsuperscript{(4)}
        & $108.19^{+0.95}_{-0.97}$ \textsuperscript{(9)}
        & $87.47^{+0.18}_{-0.19}$ \textsuperscript{(14)}
        & ---
        & --- \\

        $db/dt$ [yr$^{-1}$]
        & $-0.016 \pm 0.004$ \textsuperscript{(5)}
        & $-0.0228^{+0.0050}_{-0.0018}$ \textsuperscript{(10)}
        & $-0.0144 \pm 0.0025$ \textsuperscript{(14)}
        & $-0.0116 \pm 0.0036$ \textsuperscript{(17)}
        & $0.0173 \pm 0.0006$ \\

        Precession period [yr]
        & $\sim 500$ \textsuperscript{(5)}
        & $\sim 970$ \textsuperscript{(10)}
        & $\sim 890$ \textsuperscript{(14)}
        & Not constrained
        & $\sim 860$ \\


        Stellar companion(s)
        & Confirmed \textsuperscript{(6)}
        & Confirmed \textsuperscript{(11)}
        & Unconfirmed
        & Not found
        & Not found \\

        \noalign{\smallskip} \hline
    \end{tabular}
    \label{tab:nodal-precessiong_objects}
    }
    \tablefoot{Some values are rounded for readability.}
    \tablebib{\textsuperscript{(1)}\cite{Morton16}, \textsuperscript{(2)}\cite{Barnes11}, \textsuperscript{(3)}\cite{Esteves15}, \textsuperscript{(4)}\cite{Masuda15}, \textsuperscript{(5)}\cite{Szabo12}, \textsuperscript{(6)}\cite{Santerne12}, \textsuperscript{(7)}\cite{CCameron10}, \textsuperscript{(8)}\cite{Turner16}, \textsuperscript{(9)}\cite{Watanabe22}, \textsuperscript{(10)}\cite{Johnson2015}, \textsuperscript{(11)}\cite{Mugrauer19}, \textsuperscript{(12)}\cite{Borsa19}, \textsuperscript{(13)}\cite{Gaudi17}, \textsuperscript{(14)}\cite{Stephan22},  \textsuperscript{(15)}\cite{Cabot21}, \textsuperscript{(16)}\cite{Simonnin25}, \textsuperscript{(17)}\cite{Watanabe24}. Parameters of HD~210058 are given by TESS Input Catalogue (Table~\ref{tab:stellar_params}), parameters of HD~210058~b come from this work.}
\end{table*}

\FloatBarrier 

\clearpage
\section{Frequencies detected in BD+37~4838}

\begin{table*}[ht!]
    \caption{Frequencies identified in the full \texttt{SPOC} short-cadence data set.}
    \centering
    \begin{tabular}{c c c c}
\hline \hline
\noalign{\smallskip}
ID    &    $f \rm (d^{-1})$    &    $A_T$ (mmag)    &    $\phi$ (rad/2$\pi$)    \\ \hline
\noalign{\smallskip}
F1	& $2.3159\pm0.00001$ & $0.99\pm0.07$ & $0.036\pm0.009$\\
F2	& $2.200228\pm0.000009$ & $1.26\pm0.05$ & $0.841\pm0.003$\\
F3	& $21.143\pm0.0002$ & $0.9\pm0.4$ & $0.7\pm0.01$\\
F4	& $1.61668\pm0.000002$ & $0.9\pm0.2$ & $0.177\pm0.003$\\
F5	& $2.3868\pm0.0004$ & $0.82\pm0.15$ & $0.9\pm0.2$\\
F6	& $2.255\pm0.002$ & $0.84\pm0.07$ & $0.45\pm0.01$\\
F7	& $8.575652\pm0.000005$ & $0.64\pm0.01$ & $0.988\pm0.003$\\
F8	& $0.849726\pm0.000009$ & $0.5\pm0.2$ & $0.748\pm0.003$\\
F9	& $2.46054\pm0.00004$ & $0.52\pm0.04$ & $0.8798\pm0.0005$\\
F10	& $19.51766\pm0.00001$ & $0.32\pm0.01$ & $0.791\pm0.007$\\
F11	& $10.19775\pm0.00001$ & $0.3\pm0.2$ & $0.816\pm0.007$\\
F12	& $21.96325\pm0.00002$ & $0.29\pm0.02$ & $0.08\pm0.01$\\
F13	& $22.976\pm0.003$ & $0.29\pm0.02$ & $0.7\pm0.2$\\
F14	& $8.74243\pm0.00002$ & $0.26\pm0.01$ & $0.788\pm0.007$\\
F15	& $18.58748\pm0.00002$ & $0.19\pm0.01$ & $0.15\pm0.02$\\
F16	& $26.02688\pm0.00002$ & $0.17\pm0.02$ & $0.33\pm0.02$\\
F17	& $19.822\pm0.002$ & $0.14\pm0.03$ & $0.15\pm0.03$\\
F18	& $0.93054\pm0.00002$ & $0.29\pm0.03$ & $0.39\pm0.01$\\
F19	& $0.11581\pm0.000009$ & $0.3\pm0.1$ & $0.454\pm0.007$\\
F20	& $2.03\pm0.03$ & $0.3\pm0.1$ & $0.4\pm0.3$\\
F21	& $3.489\pm0.003$ & $0.29\pm0.03$ & $0.5\pm0.1$\\
F22	& $4.36675\pm0.00002$ & $0.28\pm0.01$ & $0.51\pm0.008$\\
F23	& $1.70704\pm0.00002$ & $0.27\pm0.02$ & $0.81\pm0.01$\\
F24	& $10.35879\pm0.00004$ & $0.11\pm0.01$ & $0.57\pm0.03$\\
F25	& $22.75497\pm0.00005$ & $0.1\pm0.2$ & $0.79\pm0.03$\\
F26	& $26.169\pm0.006$ & $0.09\pm0.03$ & $0.9\pm0.2$\\
F27	& $23.16225\pm0.00004$ & $0.09\pm0.01$ & $0.23\pm0.04$\\
\noalign{\smallskip} \hline 
    \end{tabular}
    \tablefoot{$f$ denotes oscillation frequency, $A_T$ is amplitude, $\phi$ is phase shift of the given sinusoidal component}
    \label{tab:TOI-1414Frequnencies}
\end{table*}

\clearpage
\section{The age of HD~210058 and a parent association}
\newcommand{\starname}{HD~210058}

\begin{figure}[hb]
	\begin{center}
		\includegraphics[width = 0.6\textwidth]{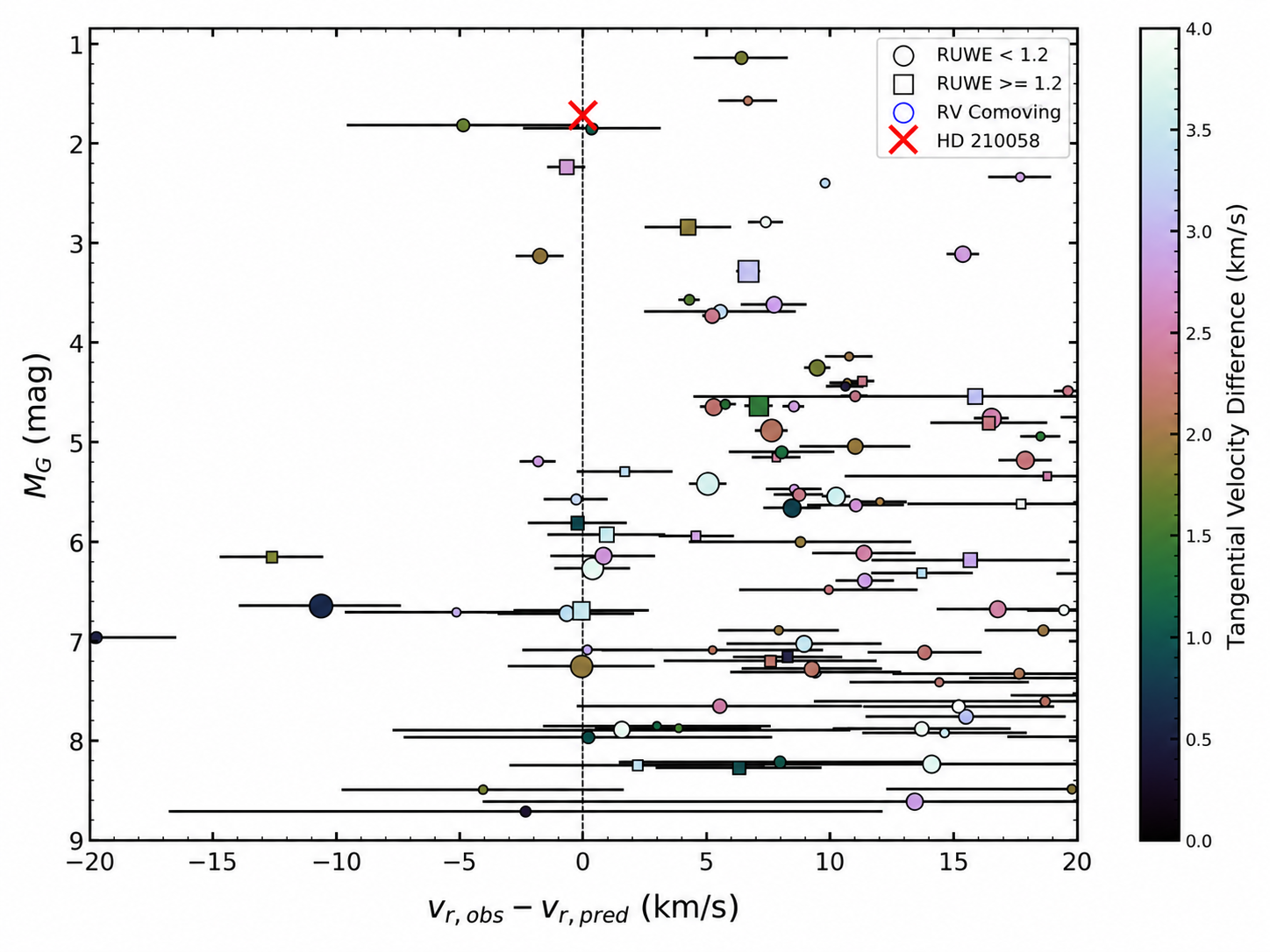}
		\caption{The difference between the predicted RV (assuming agreement with \starname) and the \textit{Gaia} RV for similar 5-dimensional positions and kinematics to \starname. This is a strong test, because RVs are not used in the initial selection (only parallax, proper motion, and position). An overdensity of points is seen around zero (perfect agreement), although the large clustering of points at +5-10~$\rm km\,s^{-1}$ off highlights how small overdensities can occur by chance.}
        \label{fig:drv}
    \end{center}
\end{figure}

\begin{figure}[th]
	\begin{center}
		\includegraphics[width = 0.6\textwidth]{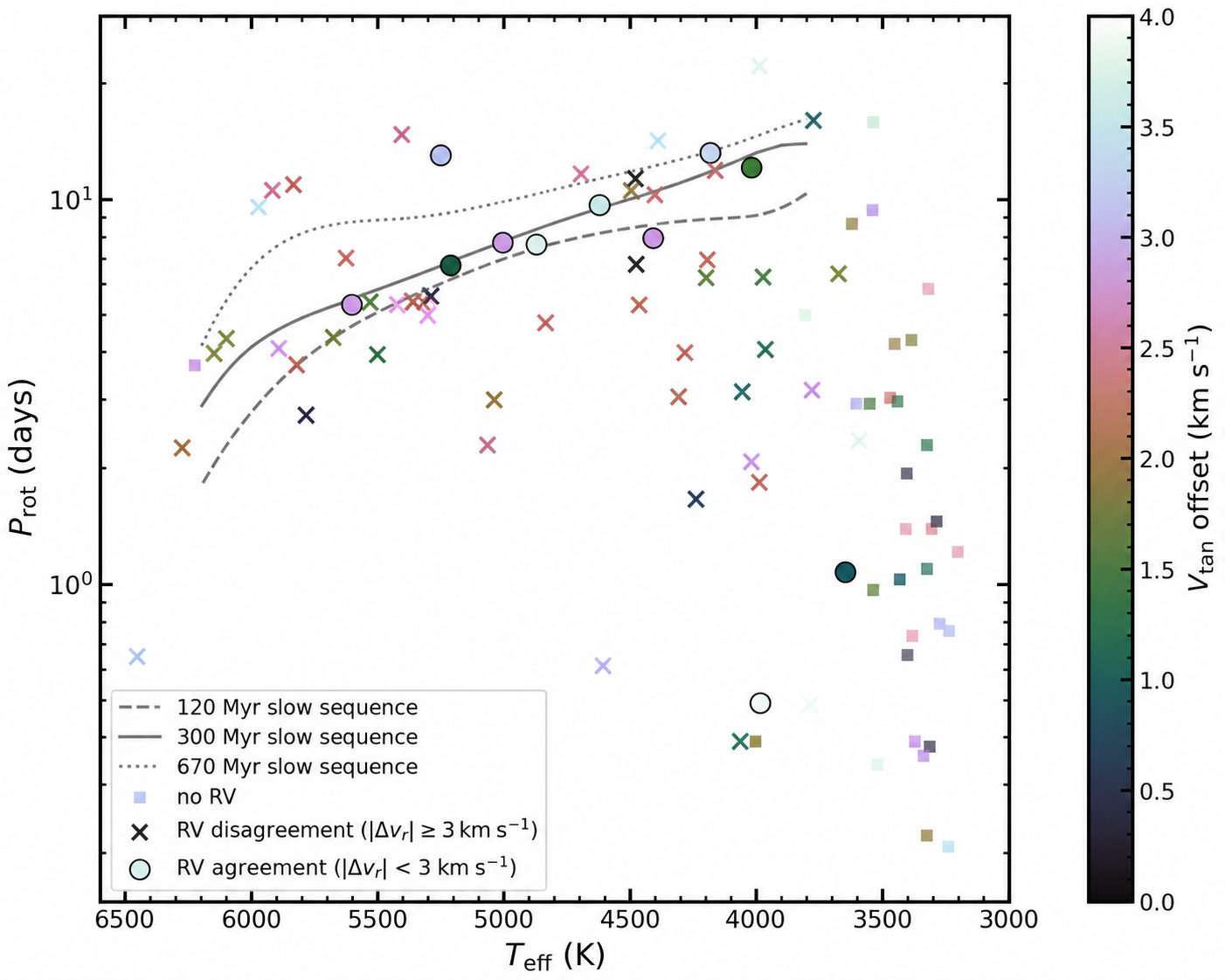}
		\caption{Rotation periods of candidate co-moving stars with \starname{} taken from the TARS catalog \citep{Boyle2026} alongside predictions from \texttt{gyro-interp} \citep{Bouma2023} drawn from the slow-rotating sequence of nearby benchmark clusters. Most stars that are RV consistent with \starname{} have rotation periods consistent with a single-age ($200-300\,\rm Myr$) population, providing additional evidence that the group is real.} 
        \label{fig:prot}
    \end{center}
\end{figure}

\end{appendix}
\end{document}